\documentclass[reprint,superscriptaddress,amsmath,amssymb,aps,prb,floatfix]{revtex4-1}
\usepackage{graphicx}
\usepackage{dcolumn}
\usepackage{bm}
\usepackage{subfigure}
\usepackage[margin=1.5cm]{geometry}
\usepackage{times}
\usepackage[T1]{fontenc}
\usepackage{amsfonts,verbatim,dsfont}
\usepackage{amssymb,amsmath,amsfonts,latexsym,graphicx,verbatim,dsfont}
\usepackage[matrix,frame,arrow]{xy}
\usepackage{epsf}
\usepackage{upgreek}
\usepackage[usenames,dvipsnames]{color}
\usepackage[english]{babel}
\usepackage{times}
\usepackage{latexsym}
\usepackage{fancyhdr}
\usepackage{setspace}
\usepackage{bbm}
\usepackage{enumerate}
\usepackage{multirow}

\usepackage{latexsym}
\usepackage{bbm}
\usepackage{enumerate}
\usepackage{listings}
\usepackage{multirow}
\usepackage[table]{xcolor}
\usepackage{algorithm}
\usepackage[bottom]{footmisc}
\usepackage[noend]{algpseudocode}
\usepackage[numbered,framed]{matlab-prettifier}
\usepackage{etoolbox}

\renewcommand{\emph}{\textit}

\newcommand{\id}{\mathds{1}}

\newcommand{\xa}{\alpha}

\newcommand{\xd}{\delta}

\newcommand{\xe}{\upvarepsilon}
\newcommand{\xg}{\gamma}
\newcommand{\xt}{\theta}

\newcommand{\xo}{\omega}

\newcommand{\xs}{\sigma}

\newcommand{\pp}{\perp}
\newcommand{\app}{\approx}

\newcommand{\Cs}{{}^{13}\R{C}}
\newcommand{\Ns}{{}^{14}\R{N}}
\newcommand{\Hs}{{}^{1}\R{H}}

\newcommand{\mR}[0]{\mathcal{R}}

\newcommand{\bn}[0]{\hat{\mathbf n}}

\newcommand{\xD}{\Delta}

\newcommand{\xT}{\Theta}

\newcommand{\mU}[0]{\mathcal{U}}

\newcommand{\bA}[0]{\mathbf A}
\newcommand{\bS}[0]{\mathbf S}
\newcommand{\bI}[0]{\mathbf I}

\newcommand{\bx}[0]{\hat{\mathbf x}}
\newcommand{\by}[0]{\hat{\mathbf y}}
\newcommand{\bz}[0]{\hat{\mathbf z}}
\newcommand{\br}[0]{\hat{\mathbf r}}

\newcommand{\fr}[2]{\frac{#1}{#2}}

\newcommand{\wt}[1]{\widetilde{#1}}

\newcommand{\sq}[1]{\sqrt{#1}}

\newcommand{\mH}[0]{\mathcal{H}}

\newcommand{\rt}{\rightarrow}

\newcommand{\e}{\exp}
\newcommand{\dg}{\dagger}

\newcommand{\beq}{\begin{equation}}
\newcommand{\eeq}{\end{equation}}
                  
\newcommand{\benum}{\begin{enumerate}}
\newcommand{\eenum}{\end{enumerate}}
                    
\newcommand{\bit}{\begin{itemize}}
\newcommand{\eit}{\end{itemize}}

\newcommand{\bea}{\begin{eqnarray}}
\newcommand{\eea}{\end{eqnarray}}

\newcommand{\bev}{\begin{verbatim}}
\newcommand{\eev}{\end{verbatim}}

\newcommand{\non}{\nonumber}

\newcommand{\zt}{\times}

\newcommand{\qt}{\tau}

\newcommand{\lb}{\left(}
\newcommand{\rb}{\right)}
\newcommand{\lsb}{\left[}
\newcommand{\rsb}{\right]}

\newcommand{\pll}{\parallel}

\newcommand{\B}[1]{\mathbf{#1}}

\newcommand{\I}[1]{\textit{#1}}
\newcommand{\R}[1]{\textrm{#1}}

\newcommand{\zl}[1]{\label{eqn:#1}}
\newcommand{\zr}[1]{Eq. (\ref{eqn:#1})}
\newcommand{\zfl}[1]{\protect\label{fig:#1}}
\newcommand{\zfr}[1]{Fig. \ref{fig:#1}}
\newcommand{\ztl}[1]{\label{table:#1}}
\newcommand{\ztr}[1]{Table \ref{table:#1}}
\newcommand{\zsl}[1]{\label{sec:#1}}
\newcommand{\zsr}[1]{Sec. \ref{sec:#1}}

\newcommand{\ket}[1]{\left\vert{#1}\right\rangle}
\newcommand{\bra}[1]{\left\langle{#1}\right\vert}

\newcommand{\ketbra}[2]{\ket{#1}\bra{#2}}

\newcommand{\expec}[1]{\left\langle #1\right\rangle}

\newcommand{\sm}{\sigma_-}

\newcommand{\ba}{\left\{ \begin{array}{lr}}
\newcommand{\ea}{\end{array}\right.}

\newcommand{\Tr}[1]{\textrm{Tr}\left\{{#1}\right\}}

\DeclareMathSymbol{\vartheta}{\mathalpha}{letters}{"12}
\DeclareMathSymbol{\theta}{\mathalpha}{letters}{"23}
\DeclareMathSymbol{\phi}{\mathalpha}{letters}{"27}
\DeclareMathSymbol{\varphi}{\mathalpha}{letters}{"1E}

\usepackage[table]{xcolor}
\newcommand{\mc}[2]{\multicolumn{#1}{|c|}{#2}}
\definecolor{Gray}{gray}{0.85}
\definecolor{LightCyan}{rgb}{0.88,1,1}
\definecolor{darksalmon}{rgb}{0.91, 0.59, 0.48}
\definecolor{maroon}{cmyk}{0,0.87,0.68,0.32}

\definecolor{mustard}{rgb}{1.0, 0.86, 0.35}
\newcolumntype{a}{>{\columncolor{Gray}}c}
\newcolumntype{b}{>{\columncolor{white}}c}

\definecolor{Ablue}{rgb}{0.96,0.24,0.00}

\definecolor{Abluetitle}{rgb}{0.,0.24,0.51}

\definecolor{orange}{rgb}{0.96,0.24,0.00}

\usepackage[colorlinks=true , citecolor=blue,urlcolor=blue]{hyperref}

\lstMakeShortInline"

\makeatletter
\def\BState{\State\hskip-\ALG@thistlm}
\makeatother

\newcommand{\beginsupplement}{%
        \setcounter{table}{0}
        \renewcommand{\thetable}{S\arabic{table}}%
        \setcounter{figure}{0}
        \renewcommand{\thefigure}{S\arabic{figure}}%
     }

\begin{document}

\title{Quantum Interpolation for High Resolution Sensing}

\author{A. Ajoy}
\email[e-mail:]{ashokaj@mit.edu}
\affiliation{Research Laboratory of Electronics and Department of Nuclear Science \& Engineering,
Massachusetts Institute of Technology, Cambridge, MA}
\author{Y. X. Liu}
\affiliation{Research Laboratory of Electronics and Department of Nuclear Science \& Engineering,
Massachusetts Institute of Technology, Cambridge, MA}
\author{K. Saha}
\affiliation{Research Laboratory of Electronics and Department of Nuclear Science \& Engineering,
Massachusetts Institute of Technology, Cambridge, MA}
\author{L. Marseglia}
\affiliation{Research Laboratory of Electronics and Department of Nuclear Science \& Engineering,
Massachusetts Institute of Technology, Cambridge, MA}
\author{J.-C. Jaskula}
\affiliation{Research Laboratory of Electronics and Department of Nuclear Science \& Engineering,
Massachusetts Institute of Technology, Cambridge, MA}
\author{U. Bissbort}
\affiliation{Research Laboratory of Electronics and Department of Nuclear Science \& Engineering,
Massachusetts Institute of Technology, Cambridge, MA}
\affiliation{Singapore University of Technology and Design, 487372 Singapore}
\author{P. Cappellaro}
\email[e-mail:]{pcappell@mit.edu}
\affiliation{Research Laboratory of Electronics and Department of Nuclear Science \& Engineering,
Massachusetts Institute of Technology, Cambridge, MA}

\begin{abstract}  
Recent advances in engineering and  control of  nanoscale quantum sensors  have opened new paradigms in precision metrology. Unfortunately, hardware restrictions often  limit the sensor performance.  In nanoscale magnetic resonance probes, for instance, finite sampling times greatly limit the achievable sensitivity and spectral resolution. 
We develop a technique for coherent quantum interpolation that can  overcome these problems. Using a quantum sensor associated with the Nitrogen Vacancy center in diamond, we experimentally demonstrate that quantum interpolation can  achieve spectroscopy of  classical magnetic fields and individual quantum spins with  orders of magnitude finer frequency resolution than conventionally possible.
Not only is quantum interpolation  an  enabling technique to extract structural and chemical information from single biomolecules, but it can be directly applied to other quantum systems for super-resolution quantum spectroscopy.
\end{abstract}

\maketitle
Precision metrology often needs to strike a compromise between signal contrast and resolution, since  the  hardware apparatus sets  limits on the precision and sampling rate at which the data can be acquired. In some cases,  classical supersampling techniques have become a standard tool to achieve a significantly higher resolution than the bare recorded data. For instance, the Hubble Space Telescope  uses classical digital image processing algorithms like variable pixel linear reconstruction, also known as Drizzle~\cite{Fruchter97}, to construct a supersampled image from multiple low resolution images captured at slightly different angles. This technique amounts to effectively interpolating to a higher number of pixels than in the native sensor. Unfortunately, this classical interpolation method would fail for signals obtained from a quantum sensor, where the information is encoded in its quantum phase\cite{Giovannetti06}. 

Quantum systems such as trapped ions~\cite{Kotler11}, superconducting qubits~\cite{Yan12,Bylander11} and spin defects~\cite{Alvarez11,Bar-Gill12} have been shown to perform as excellent spectrum analyzers and lock-in-detectors for both classical and quantum fields~\cite{Shi15, Lovchinsky16}. The technique relies on modulation of the quantum probe during the interferometric detection of an external field. This is typically achieved by a periodic sequence of $\pi$-pulses that invert the sign of the coupling of the quantum probe to the external field, leading to an effective time-dependent modulation $f(t)$ of the  field~\cite{Cywinski08, Ajoy11}. These  sequences, more frequently used for dynamical decoupling~\cite{Viola99b,DeLange10}, can be described by sharp band-pass filter functions obtained from the Fourier transform of $f(t)$. This description is at the basis of their application for precision spectroscopy, as   the filter is well approximated by modified sinc functions, $F(\omega \tau,N)\approx \frac{\sin^2(N\omega \tau)}{\sin^2(\omega \tau)}$, where $\tau$ is the time interval between $\pi$-pulses and $N$ the number of pulses. The filter bandpass is centered at $\nu=1/\tau$, its rejection (signal contrast) increases with the number of pulses $\propto N^2$, and the bandpass bandwidth (frequency resolution) decreases as $\Delta \nu=1/N\tau$, tremendously improving frequency resolution with increasing pulse numbers. Unfortunately, this high resolution can only be obtained if the experimental apparatus allows a correspondingly fine {time} sampling $\Delta\tau$,  with a precision $1/(N\nu)$. In practice, this is an extremely serious limitation since conventional hardware time sampling bounds are  quickly saturated, leading to losses in both signal contrast and spectral resolution.

\begin{figure*}[htbp]\centering
  \includegraphics[width=0.99\textwidth]{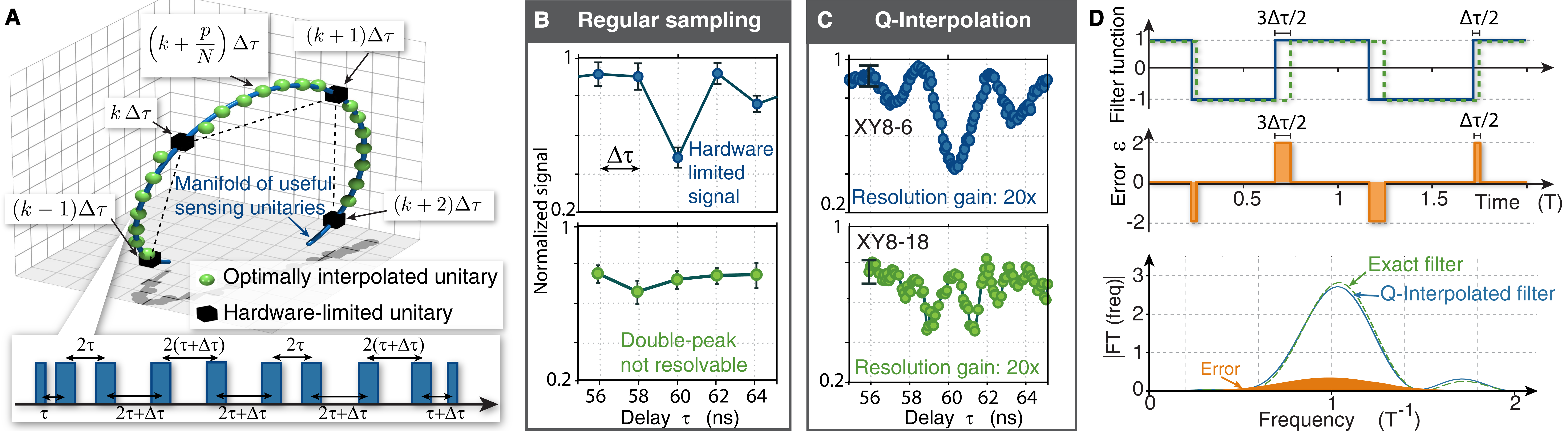}
  \caption{\textbf{Quantum Interpolation scheme. } 
  (\textbf{A}) \I{Conceptual picture of quantum interpolation}. The unitary evolution of a quantum sensor can only be probed at discrete intervals $\tau_k = k \Delta \tau$ (black cubes) and classical reconstruction would miss an accurate description (dashed black line). Quantum interpolation faithfully approximates the evolution at arbitrary small fractional intervals (green spheres) by coherent combination of pulse sequences.
  (\textbf{B-C}) \I{NMR signal from a single $^{14}$N spin associated with the NV quantum sensor.} Left:   sensing with conventional sequences limited to $\Delta \tau$=2ns. Right: quantum interpolation,  improving the resolution to 110 ps. The lower panels show that only quantum interpolation can reveal details of signal (the appearance of a double peak) linked to the quantum  evolution of the $^{14}$N spin.     (\textbf{D}) \I{Filter function description of quantum interpolation.} Top: Time domain filter function $f(t)$ for the desired (dashed green lines) and interpolated pulse sequence (solid blue lines) for the simplest case of a half-time interpolation with total sequence time $T$. The deviation between these filters is the error function $\epsilon$ (middle panel) that needs to be minimized for an optimal interpolation construction. Bottom: Frequency domain representation of both  filter functions and their difference.}
\label{fig:Scheme}
\end{figure*}

Here we introduce a technique, that we call quantum interpolation,  to overcome these limitations in sensing resolution. In analogy to classical interpolation, our method aims at capturing data points on a finer mesh than they are directly accessible because of experimental limitations.  The key idea is presented in Fig.~\ref{fig:Scheme}A: the result of any quantum sensing experiment, implemented for example, by a dynamical decoupling sequence with fixed precession time $\tau$, can be represented as a point in a {continuous} manifold of evolution operators, but the timing resolution $\Delta \tau$ limits sampling to only a discrete subset of points in this manifold. Simply acquiring data at two or more time points and interpolating the results, as done in classical sensing to obtain an approximation of the signal at  intermediate times, yields no new information. Indeed, the information is stored in the quantum sensor phase, which is then read in an incoherent manner. Instead, we achieve quantum interpolation by manipulating the quantum sensor dynamics in a coherent way, effectively  \textit{supersampling} the ideal sensing manifold at arbitrarily small fractional intervals. 
More precisely,  given discrete propagators $\mathcal U(\tau_k)$ describing the quantum probe evolution under a control sequence block of $\pi$-pulses separated by a time $\tau_k = k \Delta \tau$, we construct the interpolated propagators 
\begin{equation}
U^N(\tau_{k+ p/N}) =\mathcal P\left\{\prod_{m=1}^{N-p}\mathcal U(\tau_k)\prod_{n=1}^{p}\mathcal U(\tau_{k+1}) \right\} \ \approx \mathcal U^N(\tau_{k+ p/N}),
\label{eq:naiveinterp}	
\end{equation}
suitably ordering the pulse sequence for interpolation, reflected by the permutation $\mathcal P$.

We exploit quantum interpolation to perform high spectral resolution magnetometry of quantum and classical fields using the electronic spin of the Nitrogen Vacancy (NV) center in diamond~\cite{Taylor08} as a nanoscale probe~\cite{Bar-Gill12,Cooper14, Myers14,Lovchinsky16}. 
Using a conventional XY8-6 dynamical decoupling sequence~\cite{Staudacher13, Mamin13}  to measure the $^{14}$N nuclear spin of the NV center, we obtain a low resolution  signal where the expected narrow sinc-like dip  is barely resolved (Fig.~\ref{fig:Scheme}B). Upon increasing the number of pulses, this dip signal is completely lost. To recover the signal  with high resolution we use an optimized interpolation sequence (Fig.~\ref{fig:Scheme}C) that completely mitigates the deleterious effects of timing resolution. Indeed, the number of points that can be sampled  via quantum interpolation scales linearly with the number of pulses $N$ while the filter bandwidth decreases as $1/N$. 
The sensing resolution is now determined only by the quantum probe coherence time $T_2$ (simultaneously extended due to dynamical decoupling) and the number of pulses that can be reliably applied.

The ordering of the different pulse sequence blocks is a crucial step in achieving an interpolated propagator that would be the most faithful approximation of $\mathcal U^N(\tau_{k+p/N})$ at large $N$. For instance, a naive construction, $\mathcal P = \mathbf1$ in Eq.~(\ref{eq:naiveinterp}), leads to  error accumulation.
We tackle this problem by minimizing the deviation $\epsilon=|f_U-f_{\mathcal U}|$ of the time domain modulation (shaded regions in Fig.~\ref{fig:Scheme}C) as we find it minimizes the filter function error and maximizes the fidelity of the interpolated propagator with the ideal one. 

We can thus design a simple procedure to determine the {\textit{optimal}} control sequence to approximate any desired unitary $\mathcal U^N(\tau_{k+p/N})$. Intuitively, the optimal construction compensates the error at each decoupling sequence block and achieves a constant error for any number of pulse $N$ that depends only on $\Delta\tau$. We show analytically and numerically that the error for interpolated propagators of all $p/N$ samples is approximately equal and bounded by the error of $U^2(\tau_{k+1/2}) = \mathcal U^2(\tau_{k+1/2})+\mathcal O(\Delta \tau^2)$~(see Supplemental Information~\cite{SM}). 

\begin{figure*}[htp]
\includegraphics[width=\textwidth]{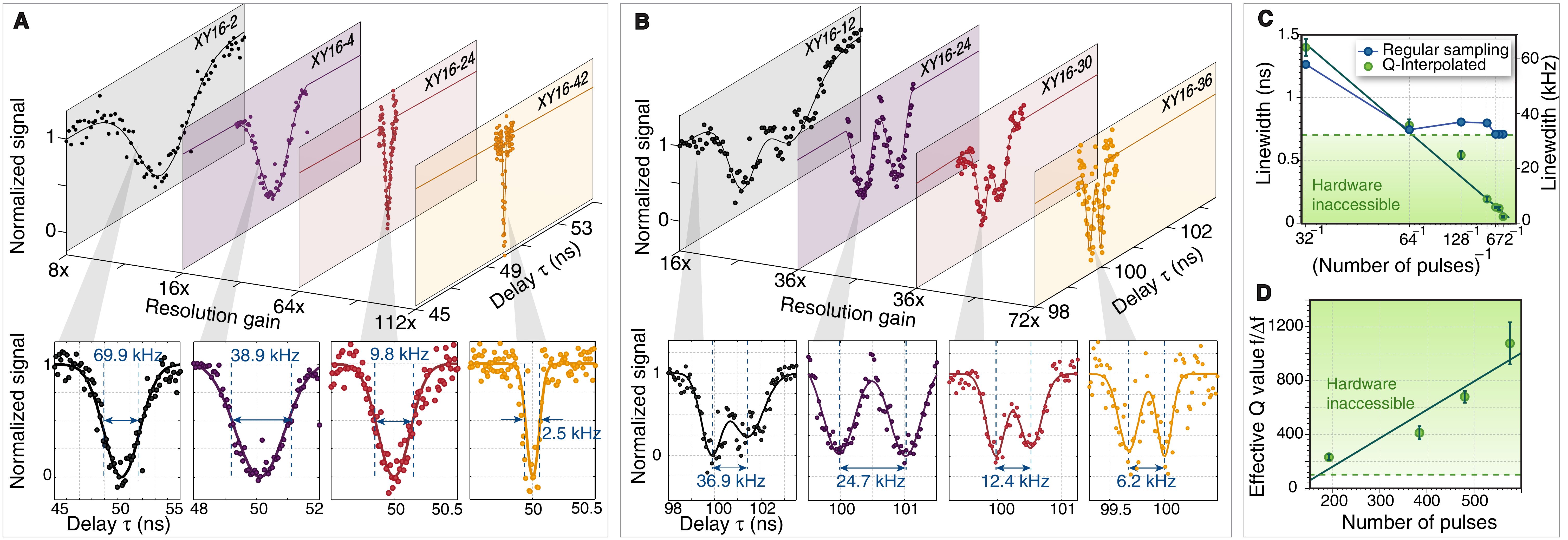}	
\caption{\footnotesize{\textbf{High resolution sensing and spectroscopy.} 
(\textbf{A}) \textit{Detection of the spurious harmonic of an AC magnetic field via quantum-interpolated XY16 sequences}. The incoherent external magnetic field is generated by an AC current at $f_{\textrm{AC}}=2.5$MHz through a 20-$\mu$m wire located in the vicinity of the NV center. Our hardware limitation ($\Delta \tau=1$ns) translates in a frequency resolution of $\Delta f_{\textrm{AC}} = 35.3$kHz, and would cause a severe suppression of the detected signal as its linewidth decreases linearly with the number of $\pi$-pulses. In the rightmost panel, quantum interpolation enables supersampling at 8.9ps (an effective boost of 112), which still permits to resolve clearly a linewidth of 2.5kHz.  
(\textbf{B}) \textit{Detection of  incoherent AC magnetic fields with two distinct frequencies.} Quantum interpolation with a maximum of 672 $\pi$-pulses allows for a gain of a factor 72 and faithfully reconstructs the AC fields, even if the two frequencies are not resolved by regular XY16 sequences with our timing resolution.
(\textbf{C}) \textit{Linewidth of the detected AC magnetometry signal} (from \textbf{A}) with regular sampling (blue) and supersampling (green). The error bars are residuals to a Gaussian fit.
(\textbf{D}) \textit{Sensing quality factor} $Q=f/\Delta f$ extracted from (\textbf{B}). Conventional dynamical decoupling sequences can only achieve $Q \leq 100$. This limit can be surpassed with quantum interpolation, scaling linearly with number of pulses, to reach $Q \approx 1000$.}}
\label{fig:AC2}
\end{figure*}

To demonstrate the power of quantum interpolation we  perform high resolution magnetometry of a classical single-tone AC magnetic field at the frequency $f_{\textrm{AC}}=2.5$MHz. By applying optimally-ordered quantum interpolated sequences (Fig.~\ref{fig:AC2}A), we detect the spurious harmonic of frequency $2f_{\textrm{AC}}$~\cite{Loretz15}.
As the number of $\pi$-pulses is increased,  the filter function associated to the equivalent XY-N sequences, and accordingly the measured signal, becomes narrower. The spectral linewidths extracted from a Gaussian fit of these dips are not affected by the finite time resolution as highlighted in Fig.~\ref{fig:AC2}C. Without quantum interpolation, we reach our experimental resolution limit after applying a sequence of only 64 $\pi$-pulses~\cite{SM}.  Quantum interpolation enables AC magnetometry far beyond this limit: we obtain an improvement by a factor 112 in timing resolution, corresponding to sampling at 8.9ps.

The advantage of quantum interpolation over conventional dynamical decoupling sequences is evident when the goal is  to resolve signals with similar frequencies.  Fig.~\ref{fig:AC2}D shows that  our quantum sensor is easily able to detect a classic dual-tone perturbation, resolving fields that are separated by $\Delta f$ = 6.2kHz, far below the limit set by our native 1ns hardware time resolution.

A useful figure of merit to characterize the resolution enhancement of quantum interpolation, in analogy to band-pass filters, is the Q-value of the sensing peak, $Q=f/\Delta f$.  
The  Q-value for conventional decoupling pulse sequences is set by the finite time resolution, $Q=1/(2f\Delta \tau)$. 
Quantum interpolation lifts this constraint, allowing $Q \approx 2N/\pi$,  limited only by the coherence time $T_2$, $N_\mathrm{max}\leq T_2/(2\tau)$. Our experiments illustrate that the effective sensing Q can be linearly boosted with the pulse number to over 1000 (Fig. \ref{fig:AC2}D). 
Given typical NV coherence time (1ms), $\pi$-pulse length (50ns) and timing resolution (1ns), an impressive gain of about $10^4$ over the hardware limits is achievable.

Even more remarkably, the coherent construction of quantum interpolation ensures that one can measure not only classical signals, but also coherent quantum systems (e.g. coupled spins~\cite{Sushkov14}) with high spectral resolution. This result is not trivial since it implies not only modulating the quantum probe, but also  effectively \textit{engineering} an interpolated Hamiltonian for the quantum probed system~\cite{Ajoy13l}. Specifically, we consider a quantum probe (the NV center) coupled to the quantum system of interest via an interaction $\mathcal H=|0\rangle\langle0|\mathcal H_0+|1\rangle\langle1|\mathcal H_1$. Here $|0\rangle$, $|1\rangle$ are the two eigenstates of the quantum probe and $\mathcal H_{0,1}$ the target system Hamiltonians in each manifold. Then, the propagator under a $\pi$-pulse train (with timings as in the CPMG~\cite{Carr54,Meiboom58} sequence) is given by $\mathcal U^N(\tau)=|0\rangle\langle0|\mathcal U_0^N(\tau)+|1\rangle\langle1|\mathcal U_1^N(\tau)$, with
\begin{equation}
\mathcal U_{0,1}^N(\tau)=(e^{-i\mathcal H_{0,1}\tau}e^{-i\mathcal H_{1,0}2\tau}e^{-i\mathcal H_{0,1}\tau})^N
\label{eq:CPMG}	
\end{equation}
Sensing of the external quantum system is achieved via interference between the two evolution paths given by $\mathcal U_{0,1}^N(\tau)$, which results in a signal $S=\left[1+\textrm{Tr}(\mathcal U_0^N\mathcal U_1^{N\dag})\right]/2$~\cite{Taminiau12,Kolkowitz12a}. The interference is enhanced  by increasing the number of pulses $N$, and by a careful choice of the time $\tau$, making one  susceptible once again to finite timing resolution. 
Quantum interpolation can overcome this limitation,  constructing any propagator $U_{0,1}^N(\tau_{k+p/N})$ by suitably combining $\mathcal U_{0,1}^{N-p}(\tau_k)$ and $\mathcal U_{0,1}^p(\tau_{k+1})$. It is somewhat surprising that such a prescription might work at all: the non-commutativity of the propagators and the non-convergence of the perturbative Baker-Campbell-Hausdorff expansion could potentially amplify the error, when considering a large number of pulses. Fortunately, the construction developed for classical fields still  keeps the error small~\cite{SM}.

\begin{figure}[b!]\centering
  \includegraphics[width=0.375\textwidth]{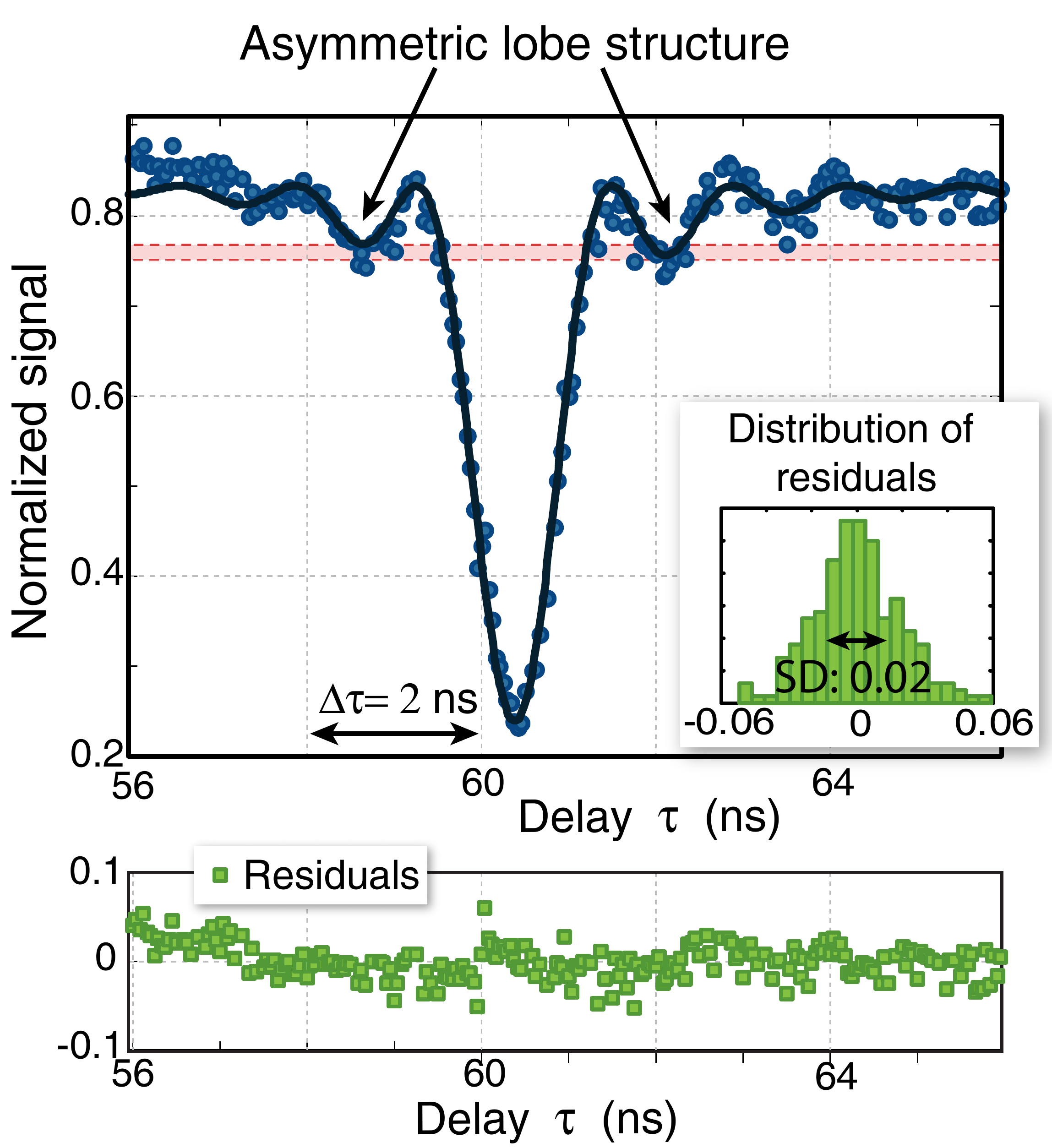}
  \caption{\textbf{High resolution spin detection.} A modified XY8-12 sequence enables an effective sampling at 48ps (a resolution gain by 41 with respect to the hardware-imposed $\Delta \tau$=2ns).  The lineshape of the $^{14}$N NMR signal displays a slight asymmetry in the signal sidelobes, an expected feature~\cite{SM} of the NMR signal under the XY8 sequence (solid line). The agreement with the theory-fitted curve is very good, reflected by the relative residual standard deviation being 3\%.}
\label{fig:Error}
\end{figure}

 Consider for example the coupling of a quantum probe (the NV center) to a two-level system (a nuclear spin-$1/2$). NV centers implanted a few nanometers below the diamond surface have recently emerged as the prime technology towards the long-standing goal of obtaining high spatial-resolution structure of single molecules in their natural environment, by performing nano-scale Nuclear Magnetic Resonance (NMR) spectroscopy~\cite{Rugar15,Lovchinsky16}.  
The outstanding key challenge is resolving the spectral features (and hence positions) of densely packed networks of spins in such molecules. 
Frequency differences, as small as a few Hz, arise from chemical shifts and the  coupling to the NV. 
The Hamiltonian of each spin in the molecule is given by $\mathcal H_{0}= \omega_{L} I_{zj}\:;\: \mathcal H_{1}=\omega_L  I_{zj} +  \sum_{\nu}A^{(j)}_{z\nu} I_{\nu j}$, where $\omega_L$ is the Larmor frequency of the spins, and $A^{(j)}_{z\nu}$ are the components of the  coupling to the NV center. Then, the $\mathcal U_{0,1}$ propagators are composed of nuclear spin rotations conditioned on the NV state; the maximum interference signal arises when  $\tau = \pi/\left[ 2(\omega_L + A^{(j)}_{zz})\right]$, the propagators corresponding to rotations around two non-parallel axes separated by an angle $\alpha=\tan^{-1} \left[ A^{(j)}_{z\perp}/(\omega_L+A^{(j)}_{zz})\right]$. The angle between the nuclear spin rotation axes in the two NV manifolds is  amplified with every subsequent application of $\pi$-pulse, giving rise to a signal contrast that grows with $N^2$. 
The destructive interference is also amplified away from the sensing peak, leading to a sinc linewidth that falls as $1/(N\tau)$, similar to the results obtained using the semi-classical filter picture.

To experimentally demonstrate the high precision sensing reached by quantum interpolation, we measure  the  $^{14}$N nuclear spin via its coupling to the NV center electronic spin. Even if the $^{14}$N is strongly coupled to the NV ($A_{zz}=-2.16$MHz), it usually does not give rise to an interferometric signal, because of its transverse coupling $A_{zx}=0$. However, a small perpendicular field $B_\perp=0.62$G generates an effective transverse coupling $\frac{\gamma_eB_\perp A_{xx}}{(\Delta - \gamma_eB_z)}$, with $A_{xx}=-2.62$MHz\cite{Chen15} and $\gamma_e=2.8$MHz/G the NV gyromagnetic ratio. 
This effect becomes sizable at a longitudinal magnetic field $B_z=955.7$G that almost compensates the NV zero-field splitting $\Delta=2.87$GHz. 
The $^{14}$N nuclear spin frequency is largely set by its quadrupolar interaction $P=-4.95$MHz, a high frequency beyond our timing  resolution (Fig.~\ref{fig:Scheme}B). We employed quantum interpolation to supersample the signal at 48ps (a 41-fold gain), revealing precise features of the spectral lineshape (Fig.~\ref{fig:Error}), including the expected slight asymmetry in sidelobes\cite{SM}. Detecting this distinct spectral feature confirms that quantum interpolation can indeed achieve a faithful measurement of the quantum signal, as we find an excellent match of the experimental data with the theoretical model, with the error being less than 3\% percent for most interpolated points. The ability to probe the exact spectral lineshape provides far more information than just the signal peaks, especially when there could be overlapping peaks or environment-broadened linewidths.

These results have immediate and far-reaching consequences for nanoscale NV-NMR~\cite{Ajoy15,Lovchinsky16,Shi15}, where our technique can map spin arrangements of a nearby single protein with a spatial resolution that dramatically improves with the number of pulses. The Q-value provides an insightful way to quantify the resolution gains for these applications. With a $Q\approx 10^4$ that is currently achievable, ${}^{13}$C chemical shifts of aldehyde and aromatic groups can now be measured~\cite{Ernst}. Beyond sensing nuclear spins, we envision quantum interpolation to have important applications in condensed matter, to sense high  frequency (hence high Q) signals~\cite{vandersar15}, such as those arising from the excitation of spin-wave modes in magnetic materials like Yittrium Iron Garnett~\cite{Serga10}.

In conclusion, we have developed a quantum interpolation technique that achieves  substantial gains in quantum  sensing resolution. We demonstrated its advantages by performing  high frequency-resolution magnetometry of both classical fields and single spins using  NV centers in diamond. The technique allows pushing spectral resolution limits to fully exploit the long coherence times of quantum probes under decoupling pulses.  We experimentally demonstrated resolution gains by 112, and Q-value by over 1000, although the ultimate limits of the technique can be at least an order of magnitude larger. Quantum interpolation thus turns quantum sensors into high-resolution and high-Q spectrum analyzers of classical and quantum fields. We expect quantum interpolation to be an enabling technique for nanoscale single molecule spectroscopy at high magnetic fields~\cite{Kost15, Hemmer13}, allowing the discrimination of chemical shifts and angstrom-resolution single molecule structure.

\I{Acknowledgments} -- It is a pleasure to thank S. Lloyd, R. Walsworth, F. Jelezko, M. Lukin, F. Casola and D. Glenn for stimulating discussions and
encouragement.  This work was supported in part by the NSF CUA and the U.S. Army Research Office.


\pagebreak
\newpage

\cleardoublepage

\newgeometry{margin=0.35in}

\setcounter{equation}{0}
\setcounter{figure}{0}
\setcounter{table}{0}
\setcounter{page}{1}
\makeatletter
\renewcommand{\theequation}{S\arabic{equation}}
\renewcommand{\thefigure}{S\arabic{figure}}
\renewcommand{\thetable}{S\arabic{table}}
\renewcommand{\thepage}{\roman{page}}
\renewcommand{\bibnumfmt}[1]{[S#1]}
\renewcommand{\citenumfont}[1]{S#1}

\pagenumbering{arabic}
\onecolumngrid
\begin{center}
\textbf{\large{\I{Supplementary Information:} Quantum Interpolation for High Resolution Sensing}}\\
\hfill \break
A. Ajoy,$^{1,\textcolor{red}{\ast}}$ Y. X. Liu,$^{1}$ K. Saha,$^{1}$ L. Marseglia,$^{1}$ J.-C. Jaskula,$^{1}$ U. Bissbort,$^{1,2}$ and P. Cappellaro,$^{1,\textcolor{red}{\dagger}}$\\
\smallskip
\emph{${}^{1}$ {\small Research Laboratory of Electronics and Department of Nuclear Science \& Engineering,
Massachusetts Institute of Technology, Cambridge, MA}}\newline
\emph{${}^{2}$ {\small Singapore University of Technology and Design, 487372 Singapore}}
\end{center}
\twocolumngrid

\beginsupplement
\tableofcontents
\section*{Summary}
In this supplement, we provide details of the theory of quantum interpolation, and supporting information for the experiments
presented in the main paper. We also provide explicit details of the construction of the optimal quantum interpolation sequences, with the hope that this might aid their adoption in other experiments. 

The Materials and Methods section presents details about NV centers and the home-built experimental setup (\zsr{setup}) as well the Hamiltonian describing the NV center coupled to a network of nuclear spins (\zsr{system}). We further provide additional information about the experiments and the fitting model, as well as additional experimental results.

The Supplementary Text provides in depth details and analysis of the quantum interpolation scheme.
We analyze in detail the  interferometric spin sensing technique (\zsr{cpmg-main}) based on the popular CPMG/XY8 pulse sequences~\cite{Carr54,Gullion90} in order to show that the signal contrast not only grows $\propto N^2$ (with $N$  the number pulses), but more importantly to obtain an insightful  and precise understanding of the signal lineshape and  linewidth. 
After quantifying in \zsr{subsampling} how finite sampling times limit the resolution and contrast, we develop the basic theory of quantum interpolation in \zsr{supersampling}, where we specify the metric that allows one to evaluate how faithful the interpolated signal is to the true sensing signal assuming there is no limitation of finite $\xD\qt$. The theory is presented in a more simple semiclassical picture of time domain filters~\cite{Cywinski09, Ajoy11}, and a more rigorous analysis of the propagators. 
In \zsr{optimal} we describe the  optimal quantum interpolation construction that minimizes errors in supersampling, providing a graphical visualization of these constructions and a simple algorithm. 
\zsr{Q-value} details the Q-value, $Q=f/\xD f$, as a figure of merit of quantum interpolation, and quantifies gains in our experiments as well as potential improvements in other experiments, especially where the signals to be sensed are of high-Q, i.e. either with narrow linewidth  or high frequency.

\section{Materials and Methods}
\subsection{Experimental setup}
\zsl{setup}
Nitrogen-vacancy (NV) centers in diamond are substitutional nitrogen atom close to a vacancy in the carbon lattice~\cite{Jelezko06}. Their electronic spins possess remarkable quantum properties that persist at room temperature. The spin state of the negatively charged NV center has an exceptionally long coherence time and its electronic level structure allows efficient, all-optical spin polarization. The level structure of NV centers is shown in Fig. 1. The NV can be optically excited by a 532 nm laser light and it emits at 637 nm. It has a zero-field splitting of 2.87 GHz between the $m_{s}=|0 \rangle$ and $m_{s}=|\pm 1\rangle$ states. A magnetic field splits the $m_{s}=|\pm1\rangle$ levels allowing selective microwave excitation of the spin transition. 

In the experiment we used NV centers that are created in an optical grade, isotopically pure diamond ($99.99 \%$ C-12, purchased from E6) via implantation and subsequent annealing. Single NV centers are addressed using a home-built confocal microscope. In the microscope, a collimated 532 nm laser (SPROUT from Lighthouse Photonics) beam is first sent through an acousto-optic modulator (AOM, Isomet Corporation, M113-aQ80L-H) for switching and then focused using an oil immersion objective (Thorlabs N100X- PFO Nikon Plan Flour 1.3NA). The sample is mounted on a 3D-piezo scanner (Npoint) to position at the microscope focus with nm precision. The fluorescence excitation light is collected by the same objective, collimated, filtered from the 532 nm beam using a dichroic (Chroma NC338988) and then focused onto a pinhole for spatial filtering. The NV center fluorescence was filtered with a 532 nm notch filter (Semrock, BLP01-594R-25) and a 594 nm long-pass filter (Semrock, BLP01-594R-25) and collected using a single-photon counting module (Perkin Elmer SPCM-AQRH-14). 

In our experiments, we generate microwave pulses to construct quantum interpolation dynamical decoupling sequences using the following hardware:
\benum[(i)]
\item Direct synthesis of the pulses using 1.25 GS/s four channel arbitrary waveform generator~\cite{Tabor} (Model WX1284C, Tabor Electronics Ltd.). This has a timing resolution of $\xD\qt=1$ns, and is employed in experiments described in Fig. 2 of the main paper.
\item By using a microwave signal generator (Stanford Research Systems SRS 384) gated by a 500 MHz PulseBlasterESR-PRO pulse generator~\cite{Pulse} from Spincore Technologies through a microwave switch (Minicircuits ZASWA-50-DR+). The PulseBlaster has a timing resolution of $\xD\qt=2$ns, and this is employed in experiments described in Fig. 1 and Fig. 3 of the main paper.
\eenum
The MW pulses are subsequently amplified using a high power amplifier (Minicircuits LZY22+). The AWG, the AOM and the single-photon counting module were gated using TTL pulses produced by the 500 MHz PulseBlaster. The static magnetic field is generated using a 1T surface magnetization permanent magnet (BX0X0X0-N52) obtained from K\&J Magnetics. The magnet assembly is mounted on a combination of motorized translation and rotation stages (Zaber TLA series) that are used to align the field to the [111] axis of the NV center. 

\begin{figure*}[htbp]
  \centering
  	{\includegraphics[width=\textwidth]{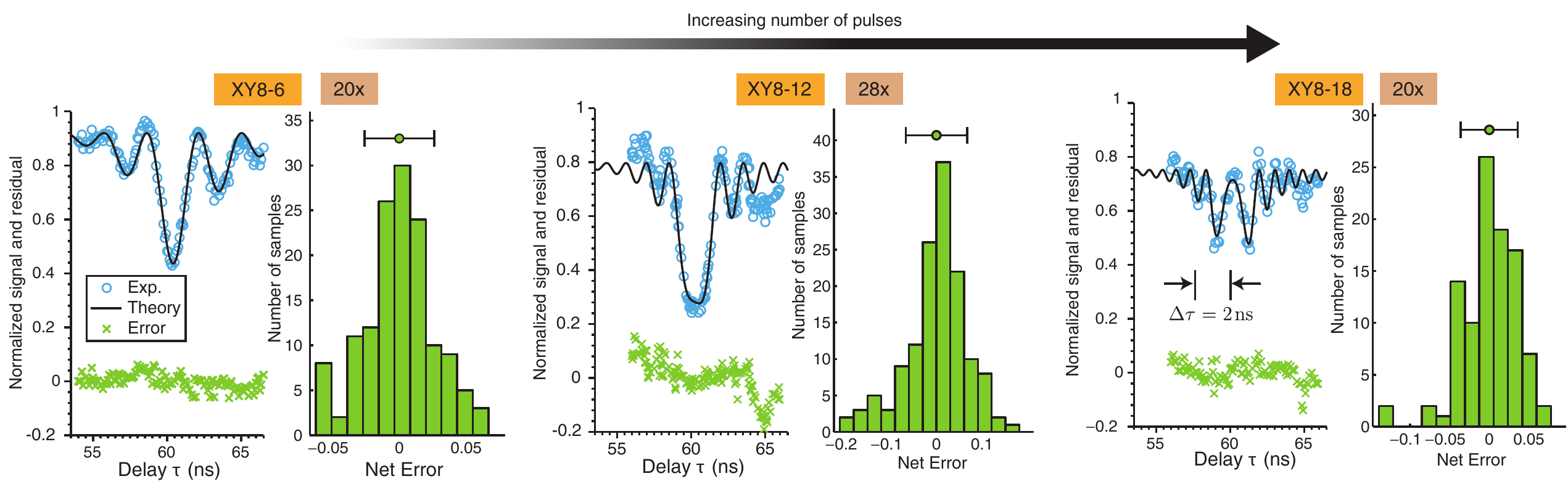}}
  \caption{\textbf{Detailed analysis of $\Ns$ spin sensing experiments.}  In these panels we study the lineshape of the signal from a $\Ns$ supersampled via quantum interpolation. This data was also depicted in Fig. 1\textbf{(B)} of the main paper. Here the experimental data (points) is fit to the expected theoretical lineshape (solid line) that is slight asymmetric. The experiments were performed at 954.71G, and from the model we extract the misalignment value of 1.14G. The crosses denote the total error of each sample from the theoretical result. The right panels show a histogram of the errors of different supersamples. The hardware resolution here was $\xD\qt=2$ns, and we effectively supersampled by the factors denoted by the x-. We sweep the number of XY8-N cycles from left to right, and the lineshape opens up characteristic sidebands upon increasing number of pulses. We find that the signal obtained via quantum interpolation is indeed a faithful representation, with the error under a few percent. }
\zfl{nitrogen-combined}
\end{figure*}

\subsection{Coupled system of NV center and nuclear spins}
\label{sec:system}
NV centers have shown to be sensitive probes of their nuclear spin environment. The NV center interacts with the nuclear spins via the anisotropic hyperfine interaction given by $$\mH_{\R{hf}} = \sum_j\bS\cdot\bA^{(j)}\cdot\bI_j= \sum_j\fr{g}{r_N^3}\lsb 3(S\cdot \br_N)(I_j\cdot \br_N) - \B{S\cdot I}\rsb,$$ where $g = \fr{\hbar\mu_0\xg_N\xg_e}{4\pi}$, with the gyromagnetic ratios of nuclear and electron spins respectively $\xg_N$ and $\xg_e$, and the vector $\vec{{\mathbf r}}_N^{(j)}=(r_{xj},r_{yj},r_{zj})$ joins the center of the NV and the nuclear spin~\cite{Childress06,Cai13b}. In the presence of a magnetic field, one can consider on the NV the pseudo two level system formed by the $\{0,-1\}$ levels. Applying now a secular approximation and retaining terms that commute with $S_z$ gives $\mH_{\R{hf}} = \sum_j\fr{g_N}{(r_N^{(j)})^3} S_z\lsb 3r_z(r_{xj}I_{xj} + r_{yj}I_{yj}) + (3r_{zj}^2 -1)I_{zj}\rsb$. The overall Hamiltonian of the coupled system is then,             
\beq
               \mH = \xD S_z +  \ketbra{0}{0}H_{\ket{0}} +\ketbra{-1}{-1}H_{\ket{-1}}
\zl{axis}
\eeq
 with $\xD=\xD_0 - \xg_eB_z$, where $\xD_0=2.87$GHz is the zero field NV splitting, and 
\begin{align}
H_{\ket{0}}&= \xo_L I_{zj}\\
H_{\ket{-1}}&=\lsb (\xo_L + A_j) I_{zj} +  B_j I_{xj} +  C_j I_{yj}\rsb
\label{eqn:NucHam}
\end{align}
represent the effective nuclear spin Hamiltonians conditioned on the state of the NV. Here, we have used the common spectroscopic notation~\cite{Abragam61}, $A_j\equiv A_{zz}^{(j)}=(3r_{zj}^2 -1), B_j\equiv A_{zx}^{(j)} = 3r_{zj}r_{xj},  C_j\equiv A_{zy}^{(j)}= 3r_{zj}r_{yj}$ to represent the magnitude of the hyperfine interactions to spin $j$, that are contained in the hyperfine tensor $\bA^{(j)} =A^{(j)}_{\mu,\nu}$.

\subsection{Experimental spin sensing via quantum interpolation}
\label{sec:expt}
In this section we provide additional information for the experiments described in the main paper, including details of the theoretical models in the fits.

\subsubsection{$\Ns$ spin sensing and lineshape analysis}

In the main paper we applied quantum interpolation based supersampling to study the lineshape from a single $\Ns$ spin intrinsic to the NV center. We performed experiments close to the ground state anti-crossing  of the NV center, $B_z\app 1000$G, where due to the presence of a weak misaligned magnetic field,  one obtains a peak signal under XY8-N of the form
\beq
S(\xd=0)=\cos (8N\xa),\;\:\R{where}\: \xa = \tan^{-1}\lsb\fr{\xg_e B_{\pp}A_{xx}}{\xD\,\xo}\rsb,
\zl{nitrogen-signal}
\eeq
with $A_{xx} = -2.62$MHz, and where $\xD = \xD_0 - \xg_eB_z$ is the resonance frequency of the NV center, and $\xo = P-A_{\pll}/2 - \xg_n B_z$, with the quadrupolar interaction $P=-4.95$MHz, the parallel hyperfine term $A_{\pll} = -2.16$MHz, and the gyromagnetic ratio $\xg_n=0.31$kHz/G. This signal originates from second order perturbation effects due to a combination of the non-secular terms $B_{\pp}S_x$ and 
$\frac{A_{\pp}}{2}(S_{+} I_- + S_- I_+)$ in the NV center Hamiltonians
\bea
 \mH&=& \mH_0 + V\\
\mH_0&=&\xD_0 S_z^2 + B_z (\gamma_e S_z + \gamma_N I_z) + P I_z^2 + A_\parallel S_z I_z\non\\
V&=&\gamma_e B_{\pp} S_x + \frac{A_\perp}{2}(S_+ I_- + S_- I_+) \non
\eea
that  yield a term $\propto S_zI_x$. The signal thus  becomes stronger close to the avoided crossing, where the energy denominator $\xD$ becomes small. For typical values of misaligned fields,$\xa$ is small, and the signal is approximately $S\propto \cos\lsb \fr{8\xg_eB_xA_{\pp}N}{\xD\:\xo}\rsb$. In Fig. 1\textbf{(B)} of the main paper, and in \zfr{nitrogen-combined}  we perform XY8 sensing while sweeping the number of cycles $N$. In \zfr{nitrogen-combined} we fit the data  to the theoretical lineshape,  numerically evaluated following \zr{signal}, 

where the operators $\mU_{\ket{0}}$ and $\mU_{\ket{1}}$ are now defined with the tilt angle $\xa_j=\xa$ from \zr{nitrogen-signal} above. 
 We find a remarkable match with the theoretical model in \zfr{nitrogen-combined}, and from the data we extract a value of $B_{\pp}$ which corresponds to an misalignment of 1.14G at the bare field of 954.71G. One is also able to discern the asymmetry in the lineshape (see \zsr{asymmetry}).
\begin{figure}[thb]
  \centering
			  {\includegraphics[width=0.45\textwidth]{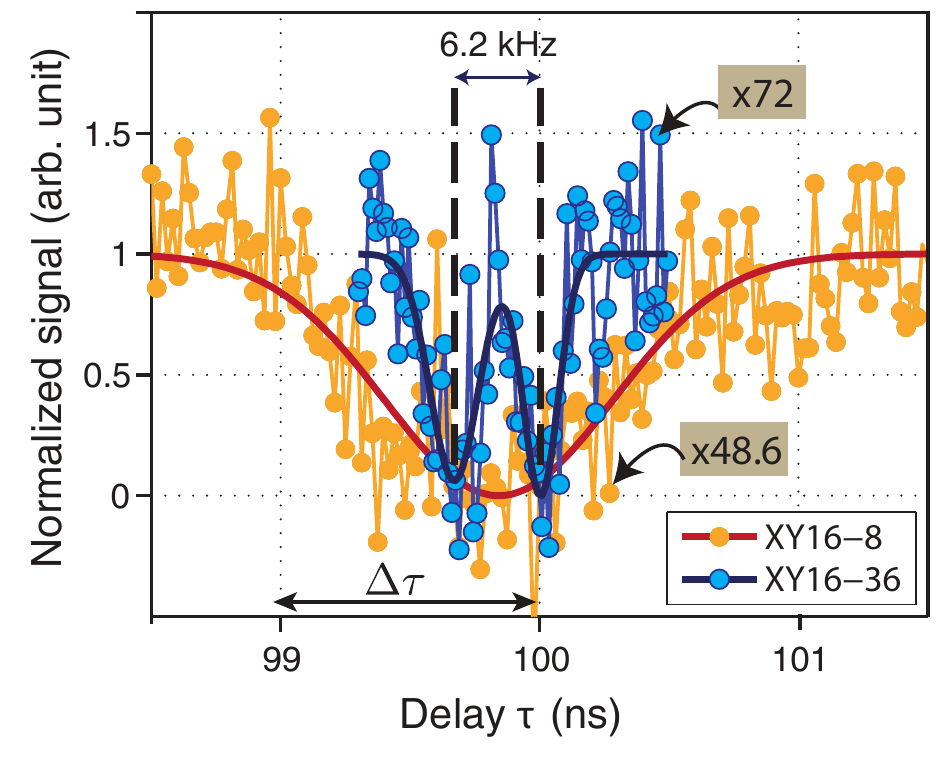}}

  \caption{\textbf{Quantum interpolation for increasing resolution.} Here we demonstrate that quantum interpolation while increasing the number of pulses can allow us to resolve peaks that were normally indistinguishable. We perform AC magnetometry from two distinct incoherent sources, separated by 6.2kHz (see also Fig. 2\textbf{(B)} of main paper). The timing resolution here experiments was $\xD\qt=1$ns, and without quantum interpolation based supersampling, the entire data would just consist of four points in this plot. The supersampling resolution gains for both experiments are indicated in the boxes.  Note that we have normalized the two experimental results so that the peak signal strength is identical for both cases. }
\zfl{double-peak-overlap}
\end{figure}
\subsubsection{Data fitting and error estimation}

\label{sec:expt-model}
To fit the theoretical model to the data,  we use a steepest descent minimization algorithm to minimize the $\chi^2$ in conjunction with simulated annealing to avoid local minima and ensure the global best fit. Subsequently, we use a Monte-Carlo approach to estimate the uncertainty of the various fit parameters. 

Let us denote the fit parameters for our model by $\mathbf P$. For a given set $\mathbf P$, our theoretical model provides a non-linear functional relation $y=f(x | \mathbf P)$. Given a measured set of data points $\{x_n\}$ and $\{y_n\}$, we determine the optimal set of parameters $\mathbf P_{\mbox{\tiny opt}}$ by minimizing $\chi^2 = \sum_n [y_n-f(x_n | \mathbf P)]^2/\sigma_y^2$. Here we have assumed that the statistical error $\sigma_y$ of the measured data points is identical for all points.

For example, the fitting parameters for Fig.~3 in the main article are the tilt angle $\alpha_1$ of the rotation axis of $U_1$, as well as the offsets and scaling factors for both the $x$ and $y$ axes ($x$ and $y$ corresponding to deviation time from the sensing peak and the measured signal intensity in this case).

Once $\mathbf P_{\mbox{\tiny opt}}$, the statistical uncertainty of $y_m$ is estimated from the deviation from the optimally fitted function $\sigma_y^2 \approx {\sum_n [y_n-f(x_n| \mathbf P_{\mbox{\tiny opt}})]^2 }/{(N-1)}$, where $N$ is the number of data points. The value of $\sigma_y$ obtained by this procedure yields sets a lower bound for the true statistical uncertainty, as any systematic deviation of the fitted function (i.e. if we have not captured the underlying true functional form in our theoretical model) increases $\sigma_y$. Subsequently the uncertainty in the fit parameters $\mathbf P$ can be estimated beyond linear order by generating artificial data sets of points $\{x_n\}$ and $\{y_n\}$ statistically distributed around $f(x_n| \mathbf P_{\mbox{\tiny opt}})$, subsequently performing a fit for each data set. We assume a Gaussian distribution for the generation of these data points, an assumption which can be verified by inspecting the distribution of $\delta y_n=y_n-f(x_n| \mathbf P_{\mbox{\tiny opt}})$ in the original data. Repeating this procedure yields a distribution of fit parameters of which the distributional form, confidence intervals and standard deviation for the individual parameters can be extracted.

\subsubsection{Spectroscopy of Classical AC Magnetic Fields}
\label{sec:expt-AC}

As a supplemental experiment to the AC magnetometry experiments described in Fig. 2\textbf{(B)} of the main paper, we performed magnetometry of two AC signals separated by 6.2kHz with XY16-8 and XY16-36 (see \zfr{double-peak-overlap}). We observe that the two peaks cannot be resolved by XY16-8, but upon increasing the number of pulses, one is able to resolve them. It is important to note that we employed quantum interpolation for both experiments; indeed given our timing resolution of $\xD\tau=1$ns, the entire data in \zfr{double-peak-overlap} would otherwise just consist of four points. 

The experiment in \zfr{double-peak-overlap}, along with those in Fig. 2 of the main paper demonstrate that via quantum interpolation, the effective ability to resolve two closeby spectral frequencies is no longer limited by hardware but only by the number of pulses that can be reliably applied.

\section{Supplementary Text}
\subsection{Interferometric spin sensing via the NV center} 
\label{sec:cpmg-main}
Although the principle of nuclear spin sensing by NV centers has been discussed extensively, the method is very often presented with a semi-classical picture of the nuclear spin noise and the filter formalism. For a better understanding of quantum interpolation, we need instead to more precisely evaluate this interferometric method by considering the full quantum mechanical evolution of the the nuclear spins~\cite{Kolkowitz12a,Taminiau12}.

\subsubsection{NV nuclear spin sensing from a geometric perspective} 
\label{sec:prop}
In the sensing pulse sequences, the NV is prepared initially in the state $\ket{\psi} = \fr{1}{\sq{2}}(\ket{0} + \ket{-1})$, while given the low magnetic field and high temperature, the nuclear spins are in the mixed state $\id_j/2$. Due to different evolutions of the nuclear spins conditioned on the $\ket{0}$ or $\ket{-1}$ of the NV center (following \zr{axis}), the evolution in the two NV manifolds gives rise to a destructive interference that is detected as a an apparent decay of the NV coherence. 

We now provide a geometric perspective to spin sensing sequences with the aim of describing the origin of the increasing sensitivity and the decreasing linewidth with the number of cycles $N$.

We first define the unitary rotation operator $\mR^j(\xT_j, \, \bn_j):=e^{-i \xT_j {\vec \sigma^j \cdot \bn_j }/{2}}$, describing a rotation of the \I{nuclear} spin $j$ around the axis $ \bn_j$ by an angle $\xT_j$ (the \I{flip} angle). 

The fundamental units of the CPMG and XY8 spin sensing sequences are described by a  unitary transformation composed of three successive rotations
\bea
\mU_{\mbox{\tiny tot}} :&=&  \mR(\xT_a, \, \bn_a ) \,\mR(\xT_b, \,  \bn_b )\, \mR(\xT_a, \, \bn_a),
\zl{three-body}
\eea
that, in turn, can be described as a rotation about a new rotation axis $\mathbf{\hat n}_{\mathrm{tot}}$ by a flip angle $\Theta_{\mathrm{tot}}$,   $\mU_{\mbox{\tiny tot}}=e^{i\phi_{\mbox{\tiny tot}} }\,\mR(\xT_{\mbox{\tiny tot}}, \bn_{\mbox{\tiny tot}})$, where $\phi_{\mbox{\tiny tot}}$ is an unimportant global phase. Some algebra yields the total effective flip angle
\bea
\xT_{\mbox{\tiny tot}}=2\arccos\left(\left|2 b \cos \lb \frac{\xT_a}{2} \rb - \cos \lb \frac{\xT_b}{2}\rb \right| \right)
\eea
and the effective rotation axis 
\begin{align}
\mathbf n_{\mbox{\tiny tot}}&=2 b \sin \lb \frac{\xT_a}{2} \rb \,  \bn_a + \sin \lb \frac{\xT_b}{2} \rb \, \bn_b \end{align}
with
\begin{align}
b &= \cos \frac{\xT_a}{2}\cos \lb \frac{\xT_b}{2} \rb - (\bn_a\cdot \bn_b) \sin \lb \frac{\xT_a}{2} \rb \sin \lb \frac{\xT_b}{2} \rb.
\end{align}

For CPMG-like sequences,  the rotation axis associated with $\mU_{\mbox{\tiny tot}}$ lies in the plane spanned by the original rotation axes $\bn_a$ and $\bn_b$, i.e. $\bn_{\mbox{\tiny tot}}$ always has the same azimuth angle as $\bn_b$ if we choose a coordinate system with $\hat{\mathbf z} = \bn_a$. We shall use this property later to visualize trajectories of metrology Hamiltonians in a three-dimensional visualization in \zfr{trajectory}. This is not the case for periodic dynamical decoupling sequences\cite{Khodjasteh07} (such as the spin echo\cite{Hahn50}).

\begin{figure}[htb]
  \centering
	{\includegraphics[width=0.3\textwidth]{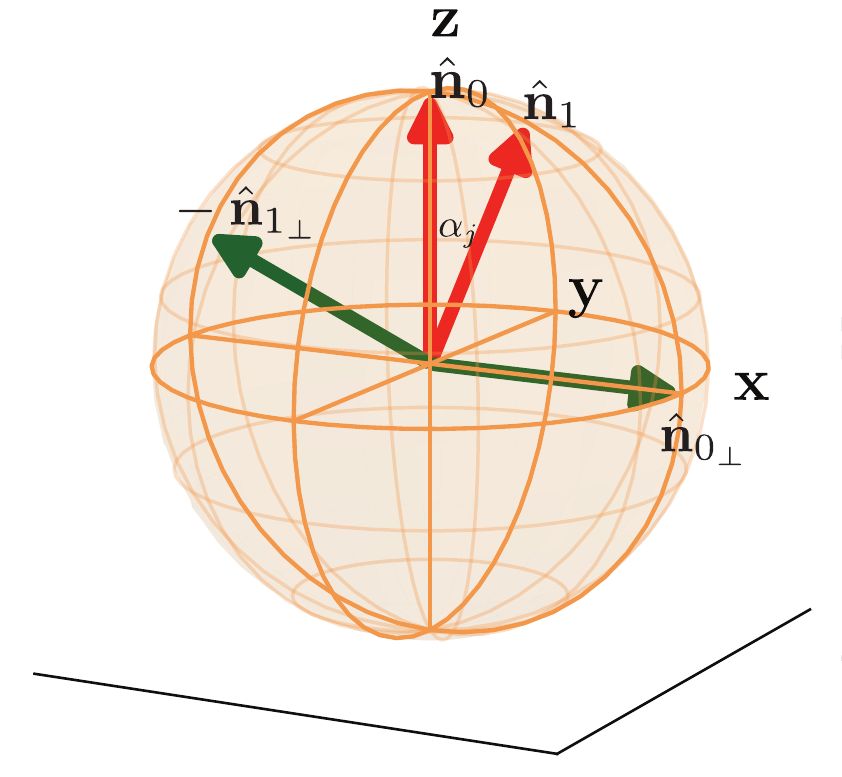}}
  \caption{\textbf{Geometry of interferometric CPMG sensing.} Bloch sphere description of the simple interferometric CPMG control sequence that is employed for sensing nuclear spins in the environment of an NV center. 
	Here $\bn_0$ and $\bn_1$ are the two axes of the nuclear spin conditioned on the state of the NV. At the signal peak, obtained for $2\qt\app \pi/\xo_L$, the result of the sequence are the two effective axes $\bn_{0_{\pp}}$ and $-\bn_{1_{\pp}}$ (see \zr{prop}).  }
\zfl{cpmg}
\end{figure}
		
We can now use these results for the system described in Sec.~\ref{sec:expt-model}, where the two axes of rotations are defined by the Hamiltonians $H_{\ket{0,1}}$ in Eq.~(\ref{eqn:NucHam}).

Here we chose the coordinate system such that $\bz=\bn_0$, i.e. the $z$-axis is aligned with the external magnetic field. We consider the coupling of the NV with a single spin $j$ at a time, which furthermore allows us to choose the coordinate system such that  $\phi_j=0$ and the hyperfine coupling $C_j=0$.
Specifically, using the geometric notation~\cite{Ajoy12b} to represent the normalized Hamiltonians in \zr{NucHam}, we have $\hat{H}_{\ket{0}} = \bn_0 = \bz$ and  $\hat{H}_{\ket{-1}}^{j} = \bn_1 = \cos(\xa_j) \bz + \sin(\xa_j) \bz_{j\pp}$, where $\bz_{j\pp} = \cos(\phi_j) \bx + \sin(\phi_j) \by$. 
We refer to the angle $\xa_j=\tan^{-1}\lsb\fr{B_j}{\xo_L + A_j} \rsb$ as the \I{tilt} angle of spin $j$.

\begin{figure}[htb]
  \centering
	{\includegraphics[width=0.45\textwidth]{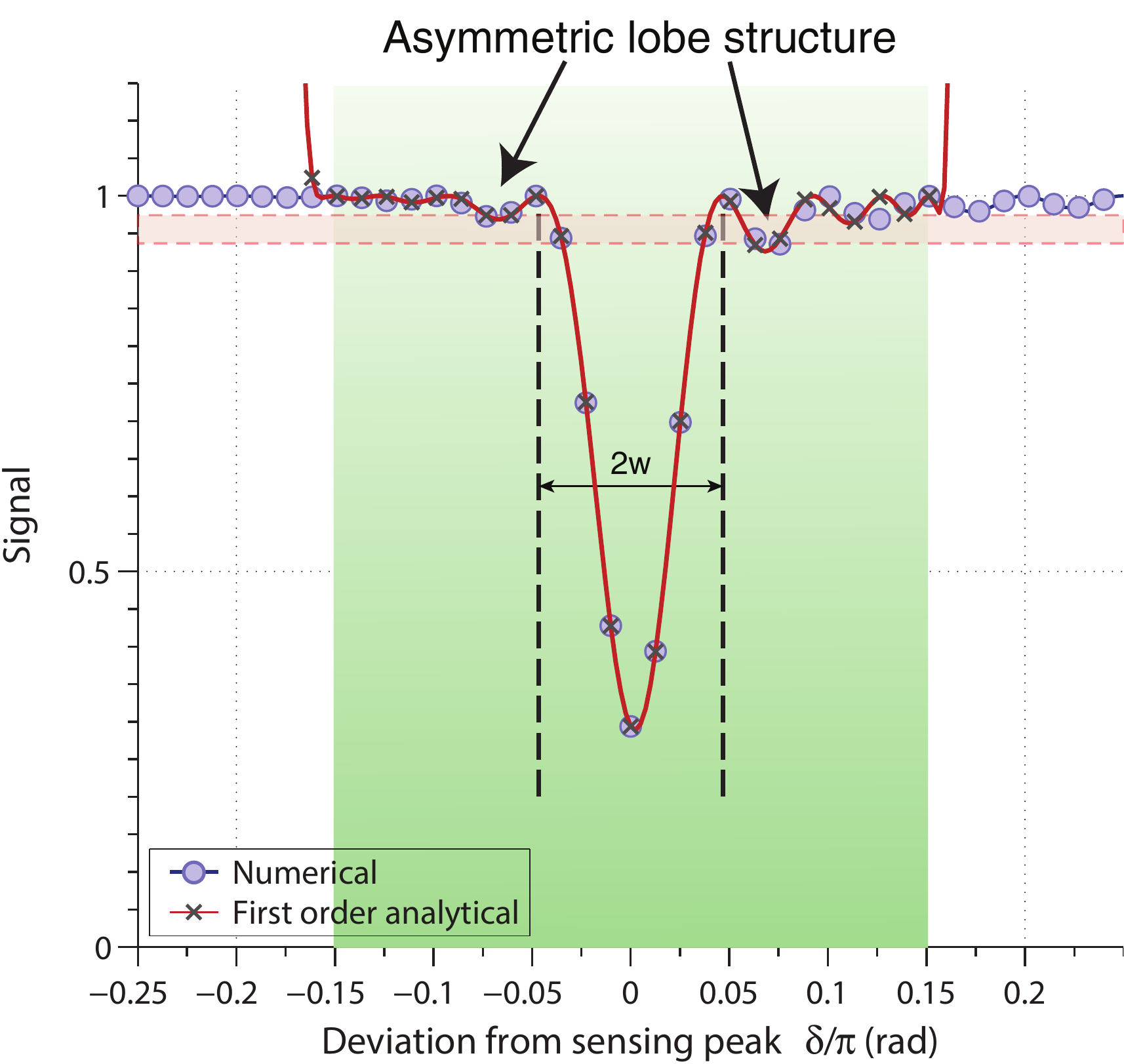}}
  \caption{\textbf{Lineshape in spin sensing experiments.} Here we compare the analytic expressions for the lineshape of spin sensing experiments with the numerically obtained result for a small deviation $\xd$ about the sensing peak. We considered $\xa_j=0.1$ rad and $N=10$ cycles of the CPMG experiment. The result demonstrates that the first order expression obtained in \zr{first-order} does indeed capture the lineshape accurately, including the asymmetry in the sidelobes on either side of the sensing peak (red dashed lines). The green shading represents the region over which the first order expression is a good approximation. The dashed lines describe the evaluation of the signal linewidth $w$ following \zr{linewidth}.}
\zfl{linewidth}
\end{figure}

The simplest protocol for spin sensing is the interferometric CPMG-N technique~\cite{Carr54, Meiboom58}, which consists of $2N$ refocusing $\pi$ pulses. For spin sensing, we sweep the delay between pulses $2\qt$, and the sensing signal dip appears when $2\qt\app \pi/\xo_L$. 
At this time, the nuclear spin sees two different evolutions conditioned on the state of the NV center, that for $\xo_L\gg A_j, B_j$ can be approximated as
\begin{align}
\mU_{\ket{0}}=\mR(\pi/2,\bn_0)\mR(\pi, \bn_1)\mR(\pi/2, \bn_0) &=& \mR(\bn_{\ket{0}},\Theta_{\ket{0}})\non\\
\mU_{\ket{-1}}=\mR(\pi/2, \bn_1)\mR(\pi, \bn_0)\mR(\pi/2, \bn_1) &=& \mR(\bn_{\ket{-1}},\Theta_{\ket{1}}).
\zl{prop}
\end{align}
(see \zsr{CPMGsignalexact} for the exact expression.) The effective axes of rotation are
\begin{align}
\bn_{\ket{0}} &= \frac{\bn_0 - \bn_1 \cos \alpha_j}{\sin \alpha_j} = -\bn_{1_{\pp}} \\
\bn_{\ket{1}} &= \frac{\bn_1 - \bn_0 \cos \alpha_j}{\sin \alpha_j}= \bn_{0_{\pp}}.
\end{align}
Note that both axes lie in the plane span($\bn_0, \, \bn_1$) and are orthogonal to $\bn_1$ and $\bn_0$ respectively. Thus they retain the same mutually spanned angle $\bn_0 \cdot \bn_1 =  \cos\alpha_j$ in magnitude.
The effective flip angles are found to be $\xT_{\ket{0}}=\xT_{\ket{-1}}= 2\xa_j$, which leads to the to a simple \I{geometric} interpretation (see \zfr{cpmg}): effectively the control protocol translates the initial tilt angle $\xa_j$ to twice the \I{flip} angle, while the effective axes are perpendicular to the initial axes are still separated by $\xa_j$.

We can now  formally derive the dip signal from a CPMG/XY8 experiment\footnote{CPMG and XY8 sequences just differ in the phases of the pulses employed, and the resulting signal in both cases is quantitatively the same. 

}, and interpret it geometrically using \zr{prop}. 
The time evolution operator for the entire control sequence with $N$ cycles is 
\begin{align}
\label{EQ:Def_U}
 U=\ket{0}\bra{0}\otimes \mU_{\ket{0}}^N + \ket{-1}\bra{-1} \otimes \mU_{\ket{-1}}^N,
\end{align}
 Note that for $\mU_{\ket{0/-1}}^N=\mR(N\Theta_{\ket{0}}, \bn_{\ket{0}} )$ the rotation angles are amplified by $N$, whereas the rotation axes remain unchanged.

Initially, we prepare the NV in the $\frac{\ket{0}+\ket{-1}}{\sqrt 2}$ state by applying a $\frac{\pi}{2}$ pulse. The initial state of the system is thus described by the density matrix $\rho_{\mbox{\tiny ini}}=\frac 1 4 (\mathbbm 1 + \sigma_x) \otimes \mathbbm 1$, where the first operator acts on the NV space and the second on the nuclear spin space. After the decoupling pulse sequence, the system is thus in the state
\beq
\label{EQ:rho_f}
\rho_{\mbox{\tiny final}} = U \rho_{\mbox{\tiny ini}} U^\dag=\frac 1 4 (\mathbbm 1 + U (\sigma_x \otimes \mathbbm 1)  U^\dag).
\eeq
After the sequence, another $\frac \pi 2 $ pulse is applied, which maps the phase onto a population of the NV state. We can therefore define the signal as the expectation value $S=\langle \sigma_x \otimes \mathbbm 1 \rangle$ before the last $\frac \pi 2 $ pulse. The signal can be interpreted as the  overlap of the initial and final density matrix
\begin{align}
S=\mbox{Tr}(\sigma_x \rho_{\mbox{\tiny final}})=4\mbox{Tr}(\rho_{\mbox{\tiny ini}}^\dag \rho_{\mbox{\tiny final}})-1.
\end{align}
Using Eq.~(\ref{EQ:Def_U}) and Eq.~(\ref{EQ:rho_f}), a straightforward calculation yields
\begin{align}
\begin{split}
\label{EQ:S_eval}
S &= \frac 1 4 \mbox{Tr}[ (\sigma_x \otimes \mathbbm 1) (  \ket{0}\bra{-1}\otimes \mU_{\ket{0}}^N {\mU_{\ket{-1}}^N}^\dag +  \ket{-1}\bra{0}\otimes \mU_{\ket{-1}}^N {\mU_{\ket{-0}}^N}^\dag     )  ]\\
&= \frac 1 4 \mbox{Tr} ( \mU_{\ket{0}}^N {\mU_{\ket{-1}}^N}^\dag + \mU_{\ket{-1}}^N {\mU_{\ket{-0}}^N}^\dag  ).
\end{split}
\end{align}
Since the trace of any SU(2) rotation operator is real, the last two terms are equal and the signal can be expressed in geometric terms 

\bea
\label{EQ:S_eval2}
S &=&  \frac 1 2 \mbox{Tr} \Big\{\left[ \mathbbm 1 \cos \lb \frac{N\Theta_{\ket{0}}}{2} \rb - i \bn_{\ket{0}} \cdot \boldsymbol \sigma \; \sin \lb \frac{N\Theta_{\ket{0}}}{2} \rb   \right]\non\\
& \times& \left[ \mathbbm 1 \cos \lb \frac{N \Theta_{\ket{-1}}}{2} \rb - i \bn_{\ket{-1}} \cdot \boldsymbol \sigma \; \sin \lb \frac{N \Theta_{\ket{-1}}}{2} \rb  \right] \Big\}\non\\
&=& 1 - \sin^2 (N\xa_j) \cos^2(\xa_j/2)
\zl{signal}
\eea
To obtain the last line, we used $(\bn_{\ket{0}} \cdot \boldsymbol \sigma )(\bn_{\ket{-1}} \cdot \boldsymbol \sigma  ) = \bn_{\ket{0}} \cdot \bn_{\ket{-1}}\mathbbm 1 + i \boldsymbol \sigma \cdot (  \bn_{\ket{0}} \times \bn_{\ket{-1}} )$ and $\mbox{Tr}(\sigma_j)=0$ for all terms containing a single Pauli matrix.

Geometrically, the signal in \zr{signal} is just the overlap of the rotations $\mU_{\ket{0}}^N=\mR(N\xa_j,-\bn_{1_\pp})$ and $\mU_{\ket{-1}}^N=\mR(N\xa_j,\bn_{0_\pp})$. It also becomes evident that amplification of the flip angle from $\xa_j$ to $N\xa_j$ upon application of $N$ cycles explains why the peak signal intensity grows quadratically with the number of cycles $N$ -- an important feature for external spin sensing. Equivalently, the application of $N$ cycles leads to a longer evolution path length and hence larger phase accumulation in the interferometric detection.

\subsubsection{Exact analysis of the signal dip}
\label{sec:CPMGsignalexact}
To evaluate the exact expression for the  peak signal from a CPMG/XY8 experiment we consider the propagators $\mU_{\ket{0}} = \mR(\eta\pi/2, \bn_1)\mR(\pi, \bn_0)\mR(\eta\pi/2, \bn_1)$, and $\mU_{\ket{-1}}=\mR(\pi/2, \bn_0)\mR(\eta\pi, \bn_1)\mR(\pi/2, \bn_0)$, where $\eta = \lsb \lb 1 + \fr{A_j}{\xo_L}\rb^2 + \lb \fr{B_j}{\xo_L}\rb^2 \rsb^{1/2}$ takes into account that the nuclear spin Hamiltonian norm in the two NV manifolds is different.

 This gives for $N$ cycles of the experiment,
\bea
\mU^N_{\ket{0}} &=& -\id\cos N\upalpha_j   +i\fr{\sin N\upalpha_j}{\sin\upalpha_j}  \boldsymbol \sigma \cdot\lsb -\bn_{1_{\pp}} \sin\xa_j \right. \\
&& \left. + \bn_1\cos\xa_j\cos(\eta\pi/2)\rsb\non\\
\mU^N_{\ket{-1}} &=& -\id\cos  N\upalpha_j  +i\fr{\sin N\upalpha_j}{\sin\upalpha_j} \boldsymbol \sigma \cdot\lsb \bn_{0_{\pp}} \sin\xa_j\sin(\eta\pi/2) \right.\non\\
&& \left. + \bn_0\cos(\eta\pi/2) \rsb\non,
\eea
where 
 $\cos(\upalpha_j) = \cos(\xa_j)\sin (\eta\pi/2)$. Note that when the interactions are weak, $\xo_L\gg A_j, B_j$, we have $\eta\rt 1$ and $\cos\upalpha_j \rt \cos\xa_j$, and one exactly recovers the expressions \zr{prop} above. The exact signal including the hyperfine terms is now,
\bea
  1-S &=& \cos^2 N\upalpha_j + \fr{\sin^2 N\upalpha_j}{\sin^2\upalpha_j}\lsb \sin^2\xa_j\cos^2\upalpha_j\lb \cos(\eta\pi/2) - 1\rb \right. \non\\
&+& \left. \cos^2\xa_j-  \cos^2\upalpha_j + \sin^2 \xa_j\cos(\eta\pi/2) \rsb\non
\eea

\begin{figure}[t]
  \centering
	{\includegraphics[width=0.48\textwidth]{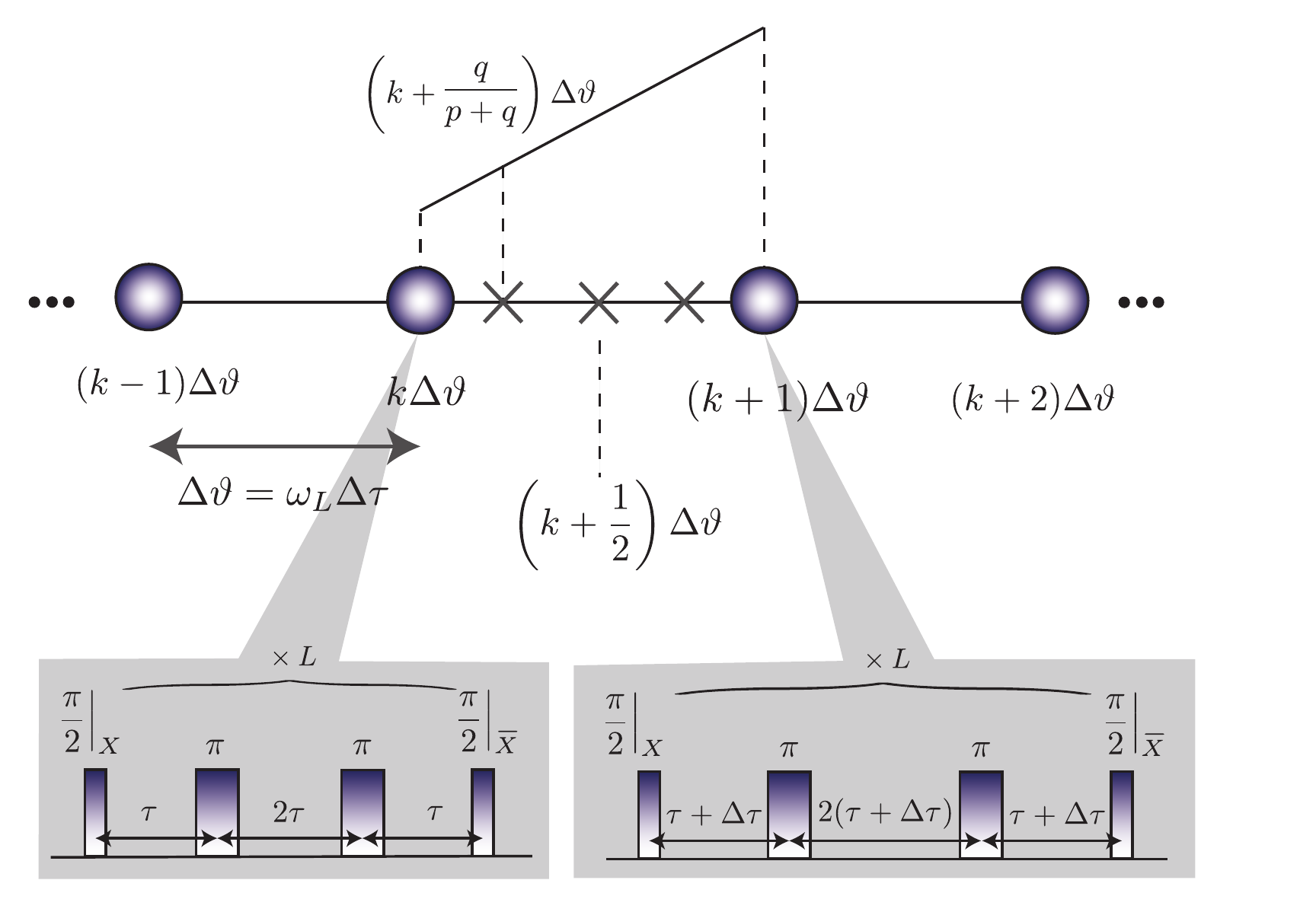}}
  \caption{\textbf{Schematic of quantum interpolation based supersampling.} Consider that we have, due to hardware limitations, a sampling resolution $\xD\qt$ in the delays $\qt$ that are swept in the usual CPMG sequence. If $\xD\xt = \xo_L\xD\qt$ this refers to the fact that one can only sample angles $k\xD\xt$, with integral $k$ (purple points). The aim of quantum interpolation based supersampling is to effectively obtain a large number of samples $\lb k + \fr{q}{p+q}\rb\xD\xt$ (crosses) between two neighboring hardware allowed samples. In principle for $N$ cycles of a CPMG sequence, the number of experimentally achievable samples with low error scales $\propto N$, allowing the effective sensing resolution to be limited only by the number of pulses that can be reliably applied. }
\zfl{supersample_panel}
\end{figure}

\subsubsection{Linewidth of the nuclear spin sensing signal}
\label{sec:asymmetry}
The previous sections considered the peak signal obtained as a result of the spin sensing experiment. However, it is also of critical importance to quantify the linewidth of the sensing signal since by falling as $1/N$ it allows sensing spins at higher resolution as $N$ increases. 

To derive the sensing linewidth, we resort to an expansion in the deviation $\xd$ about the signal peak obtained in \zr{prop},
\begin{align}
\mU_{\ket{0}}^N&(\pi + \xd)=\lsb\mR(\pi/2 + \fr{\xd}{2},\bn_0)\mR(\pi+\xd,\bn_1)\mR(\pi/2+ \fr{\xd}{2},\bn_0)\rsb^N\zl{pre-comp}\\
&\app \id\cos N\xa'_j + i\fr{\sin N\xa'_j}{\sin\xa'_j}\boldsymbol \sigma \cdot \lsb -\bn_{1_\pp}\sin\xa_j- \xd(1+\cos\xa_j)\bn_1\rsb\non
\end{align}
where to first order in $\xd$,
\beq
\sin^2\xa'_j = \sin^2\xa_j + \xd^2(1+\cos\xa_j)^2\:.
\zl{alpha-prime}
\eeq
which incorporates an effective destructive interference in the flip angle. 
It is also instructive to compare \zr{pre-comp} with \zr{prop}: the expressions are identical except for a corruption factor proportional to $\xd$ in \zr{pre-comp}. This can be visualized as a slight mixing of the perfect vector $\bn_{1_{\pp}}$ with a term $\xd(1+\cos\xa_j)\bn_1$. Crucially this is the same factor that causes the interference in \zr{alpha-prime}.  Similarly in the $\ket{-1}$ manifold of the NV center one has,
\begin{align}
\mU_{\ket{-1}}^N&(\pi + \xd)=\lsb\mR(\bn_1,\pi/2 + \fr{\xd}{2})\mR(\bn_0,\pi + \xd)\mR(\bn_1,\pi/2+ \fr{\xd}{2})\rsb^N\zl{pre-comp2}\\
&\app \id\cos N\xa'_j + i\fr{\sin N\xa'_j}{\sin\xa'_j}\boldsymbol\xs\cdot \lsb \bn_{0_\pp}\sin\xa_j - \xd(1+\cos\xa_j)\bn_0\rsb\non
\end{align}
This gives the signal similar to \zr{signal}, but now as a function of the deviation from the sensing peak $\xd$, 
\bea
S &=& \cos^2N\xa'_j + \fr{\sin^2N\xa'_j}{\sin^2\xa'_j}\lsb -\sin^2\xa_j\:\cos\xa_j \right. \non\\ &+& \xd^2\cos\xa_j\:(1+\cos\xa_j)^2-\left.2\xd\sin^2\xa_j\:(1+\cos\xa_j)\rsb.
\zl{first-order}
\eea
Figure \ref{fig:linewidth} compares the analytical expression in \zr{first-order} to an exact numerical calculation. It is evident that for most of the region close to the sensing peak (shaded region), the agreement is very close.

Importantly then the insight offered by \zr{pre-comp} allows one to intuitively understand the origin of the sensing linewidth: with increasing $\delta$, there is destructive interference of the flip-angle $\xa_j$ to $\xa'_j$ (\zr{alpha-prime}). As the number of cycles is increased, the  destructive interference effect is magnified by $N$ (\zr{first-order}) and leads to a decreasing linewidth $\propto 1/N$. 

To quantify the linewidth $w$ exactly, let us define it as the first \I{zero} of sensing signal $S$ in \zr{first-order}. This happens when the function $\sin(N\xa'_j)$ vanishes, i.e. $\xa'_j = \pi/N$, giving the linewidth in units of \I{angle},
\beq
w^2 \app \fr{\sin^2(\pi/N) - \sin^2\xa_j}{2\cos^2(\xa_j/2)}
\zl{linewidth}
\eeq
Similarly, the sensing linewidth in units of time can be evaluated as $w/\xo_L$, giving for small $\xa_j$, 
\beq
w/\xo_L\app 1/(\sq{2}\xo_L\cos(\xa_j/2))\cdot \sin(\pi/N)
\zl{linewidth-time}
\eeq that indeed falls as $\propto 1/N$ as we would expect for interferometric detection.

The linewidth  directly shows the origin of the \I{asymmetry} of the sensing peak. This is  subtle feature, characteristic of CPMG-like sequences (but not of period sequences) that we are able to discern  clearly in our experiments via quantum interpolation (Fig. 1 of main paper and \zfr{nitrogen-combined}). This shows that our quantum interpolation expansion is indeed of low error and faithfully represents the true signal.

The asymmetry is manifested by the linear term in $\xd$ in \zr{first-order}, or equivalently the odd $\sin\xd$ term in \zr{first-order}, as it is evident in \zfr{linewidth}. Indeed, the time $2\tau=\pi/\omega_L$ is not the exact signal minimum; instead, at this time, the effective vectors $[\bn_{0_\pp}\sin(\xa) - \xd(1+\cos\xa)\bn_0]$ and $[-\bn_{1_\pp}\sin\xa - \xd(1+\cos\xa)\bn_1]$ in \zr{pre-comp} and \zr{pre-comp2} are not exactly perpendicular to each other away from the sensing peak.

\subsection{Deleterious effects of finite sampling}
\label{sec:subsampling}
\subsubsection{Loss in sensing contrast}
Often we are interested in resolving spins that are very close together in frequency, for instance to be able to reconstruct their positions and the structure of the spin network of which they are part. The differences in frequency arising for instance from chemical shifts could be as small as $10^{-6}\xo_L$. In a real experimental scenario however, the rotations on the nuclear spins via the NV are effectively achieved through delayed evolution as in \zfr{cpmg}, and the construction of \zr{first-order} is prone to finite-sampling effects, leading to a loss of signal contrast and resolution. In this section, we quantify these deleterious effects in detail.

 Consider for interferometric spin sensing, we would like ideally to construct the CPMG sequence by matching the delay $2\qt = \pi/\xo_L$; however given a finite sampling resolution $\xD\qt$, one has a finite error that directly translate to a deviation from the ideal signal peak. For instance in \zr{prop}, this translates to errors in the rotation flip angles of $\pi/2$ and $\pi$ that constitute a perfect CPMG spin sensing sequence -- instead, these angles can now only be achieved to within the sampling interval $\xD\xt= \xo_L\xD\qt$ (see \zfr{supersample_panel}).

In the following, we shall quantify the deleterious effects of the this finite timing resolution: 
\begin{enumerate}[(i)]
\item Due to the fact that the signal linewidth decreases with the number of cycles $N$, finite sampling resolution $\xD\xt$ might cause the sensing peak to be \I{lost} beyond a threshold) $N_{\R{max}}$. This is experimentally demonstrated for instance in the left panels of Fig. 1\textbf{(B)} of the main paper,  where the sensing peak is just a single point or less and is not efficiently sampled.
\item For a deviation away $\xd_0$ from the perfect interferometric construction, we will show below that the signal falls away quadratically with $\xd_0$ and the number of cycles $N$. This leads, very quickly, to the underestimation of the sensing peak contrast, that can lead to significant error in reconstructing the hyperfine term $B_j$ for spin sensing experiments.
\item Finite resolution also leads to a decrease in the  maximum achievable peak signal, directly affecting the \I{sensitivity} of the NV based spin sensor.
\end{enumerate}

\begin{figure}[bht]
  \centering
	\includegraphics[width=0.5\textwidth]{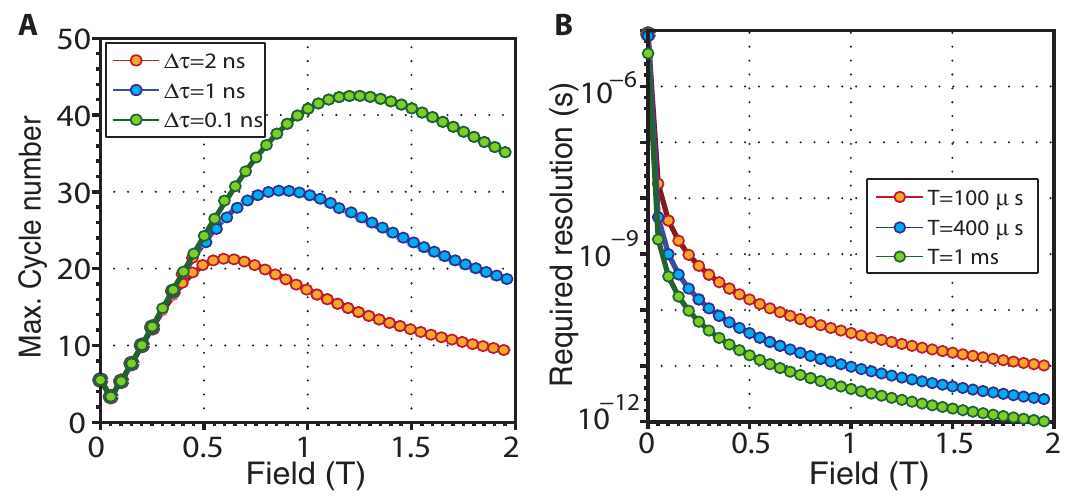}
	\caption{ \textbf{Effects of finite timing resolution.} \textbf{(A)} Panel denotes the maximum number of CPMG cycles $N$ before one becomes sensitive to finite timing resolution effects. Here we consider a single $\Hs$ spin that is 2.45nm away from the NV center at relative coordinates [1,1,2]nm to the NV center. As is evident, at moderately high magnetic fields $>0.5$T, one quickly becomes prone to subsampling effects.  \textbf{(B)} Panel denotes the largest $\xD\qt$ required to still be not prone to subsampling effects, i.e. the largest $\xD\qt$ required to still efficiently sample the signal peak. We consider here different total times of the sequence, limited by the maximum $T_2\app 1$ms. Even for moderately high magnetic fields, one requires a timing resolution of a few picoseconds, which is at the limit of current hardware (see \ztr{table}). Quantum interpolation based supersampling allows us to achieve a small effective $\xD\qt$ from modest available hardware. }
\zfl{numerical-comparison}
\end{figure}

 Let us first evaluate the maximum number of cycles $N_{\R{max}}$ such that the linewidth $w/\xo_L\leq \xD\xt$, i.e. after which we become susceptible to finite sampling effects. From \zr{linewidth},
\beq
N_{\R{max}} \app \fr{\pi}{\sqrt{(\xo_L\xD\qt)^2(1+\cos\xa_j)^2 + \sin^2\xa_j}}
\eeq
For instance for a hardware set timing limitation of $\xD\qt$=1ns (see \zfr{numerical-comparison}\textbf{(A)}), for a weakly coupled $\Hs$ spin at 0.5T and $\xa_j=0.05$rad, we have that the maximum XY8-N experiment that can be applied is $N_{\R{max}}\app 12$. This is a very small number of cycles, and increasing $N$ beyond $N_{\R{max}}$ leads to subsampling of the peak signal, leading to a substantial loss of contrast. 

Let us now in determine in detail the loss in signal contrast and resolution. Let us define sampling error $\xd_0 = \pi-k\xD\xt$ where $k$ is integral (\zfr{supersample_panel}), and which denotes the deviation from the perfect CPMG sensing sequence (the perfect sequence in \zr{prop} refers to $\xd_0=0$). The signal contrast $C(\xd_0)=\fr{1}{2}[1-S(\xd_0)]$ is now,
\beq
C(\xd_0) = \fr{1}{2}\sin^2N\xa'_j\:(1-\cos\xa'_j) + \fr{\sin^2N\xa'_j}{\sin^2\xa'_j}\sin^2\xa_j\cos\xa_j
\eeq
while the perfect contrast $C(0) = \sin^2(N\xa_j)[1+\cos(\xa_j)]$. For small sampling error $\xd_0$, one can now evaluate the effective loss in contrast,
\beq
\xe = C(0)-C(\xd_0) = \fr{1}{4}(N\xa_j)^2\xd_0^2\lb 2 - \fr{\xa_j^2}{2}\rb^2
\zl{sampling-error}
\eeq
This expression is good upto second order in $\xd_0$, and captures the scaling of the contrast loss $\xe\propto N^2\xd_0^2$, i.e. as the number of cycles $N$ increases or as one improperly samples the signal peak (larger $\xD\qt$), the loss in contrast increases quadratically. This is also evident in the experimental data shown in the left panels of Fig. 1\textbf{(B)} of the main paper -- the sensing peak is improperly sampled, and the structure in the peaks cannot be resolved.

While \zr{sampling-error} considered the loss in contrast for small $\xd_0$, let us consider now the maximum bound on the contrast $C_j(\xd_0)$. We will show that the signal not only grows quadratically slowly following \zr{sampling-error}, but is also upper bounded to a significantly lower level. Consider that maximum contrast at the peak $\left. C(0)\right|_{\R{max}} = 1+\cos\xa_j$, while at finite $\xd_0$ we have,
\beq
\left. C(\xd_0)\right|_{\R{max}} = 1+\fr{\cos(\xa_j)\lsb \sin^2(\xa_j) - \xd_0^2(1+\cos\xa_j)^2\rsb}{\sin^2(\xa_j) + \xd_0^2(1+\cos\xa_j)^2}
\zl{sampling-error2}
\eeq
It is quite easy to see that $\left. C(\xd_0)\right|_{\R{max}}< \left. C(0)\right|_{\R{max}}$. For instance, for $\xd_0\gg \sin\xa_j$ (meaning one is away from the sensing peak), we have $\left. C(\xd_0)\right|_{\R{max}} = 1-\cos(\xa_j)\rt 0$ since $\xa_j$ by definition is small.  This quantifies the intuition of destructive interference affecting the flip angle $\xa_j$ into $\xa_j'$ (see \zr{alpha-prime}) as one moves away from the sensing peak. In contrast in the perfect case $\left. C(0)\right|_{\R{max}} = \fr{1}{2}(1+\cos\xa_j)\rt 1$. Hence, in summary, one can quantify the deleterious effects of limited timing resolution $\xD\qt$ with regards to signal contrast: not only does the signal grow quadratically slower with $N$ and $\xd_0$, but it is also upper bounded to a lower level. 

\subsubsection{Loss in sensing resolution}
\label{sec:subsampling2}
In addition to a loss of signal contrast, in this section we show that finite timing resolution $\xD\qt$ also leads to a loss of sensing \I{resolution}. Consider that the effective linewidth in time units (\zr{linewidth}) is given by $\xD\qt_{\R{lw}} = w/\xo_L = \fr{[\sin^2(\pi/N) - \sin^2(\xa_j)]^{1/2}}{2\xo_L\cos(\xa_j/2)}$, however the hardware limits us to effectively a resolution of $\xD\qt$. In order to resolve the signal peak faithfully we have the requirement that $\xD\qt\leq \fr{1}{2}\qt_{\R{lw}}$. Along with the fact that the number of pulses is bounded by the coherence time, $N_{\R{max}} = \fr{T_2\xo_L}{2\pi}$, this translates to
\beq
\xD\qt \lesssim \fr{1}{2\sq{2}\xo_L\cos(\xa_j/2)}\sin \lb\fr{2\pi^2}{T_2\xo_L}\rb
\zl{subsampling3}
\eeq

\zr{subsampling3} quantifies the  fact that one needs a better timing resolution (smaller $\xD\qt$) as one goes to higher magnetic fields, or higher number of cycles. For instance (see \zfr{numerical-comparison}\textbf{(B)}), for a $\Hs$ nuclear spin at a field of 0.5T, assuming $T_2=1$ms and $\xa_j=0.05$rad typical for a weakly coupled spin, \zr{subsampling3} sets the requirement $\xD\qt\leq 15.83$ps, which is a very small required timing resolution (\ztr{table}). If $\xD\qt$ does not satisfy \zr{subsampling3}, then the sensing peak can be completely lost.  This is demonstrated also in the left panels of Fig. 1\textbf{(B)} of the main paper, where poor sampling resolution does not allow us to resolve the structure in the $\Ns$ signal (that follows \zr{first-order}).

\begin{figure*}[htb]
  \centering
  {\includegraphics[width=1\textwidth]{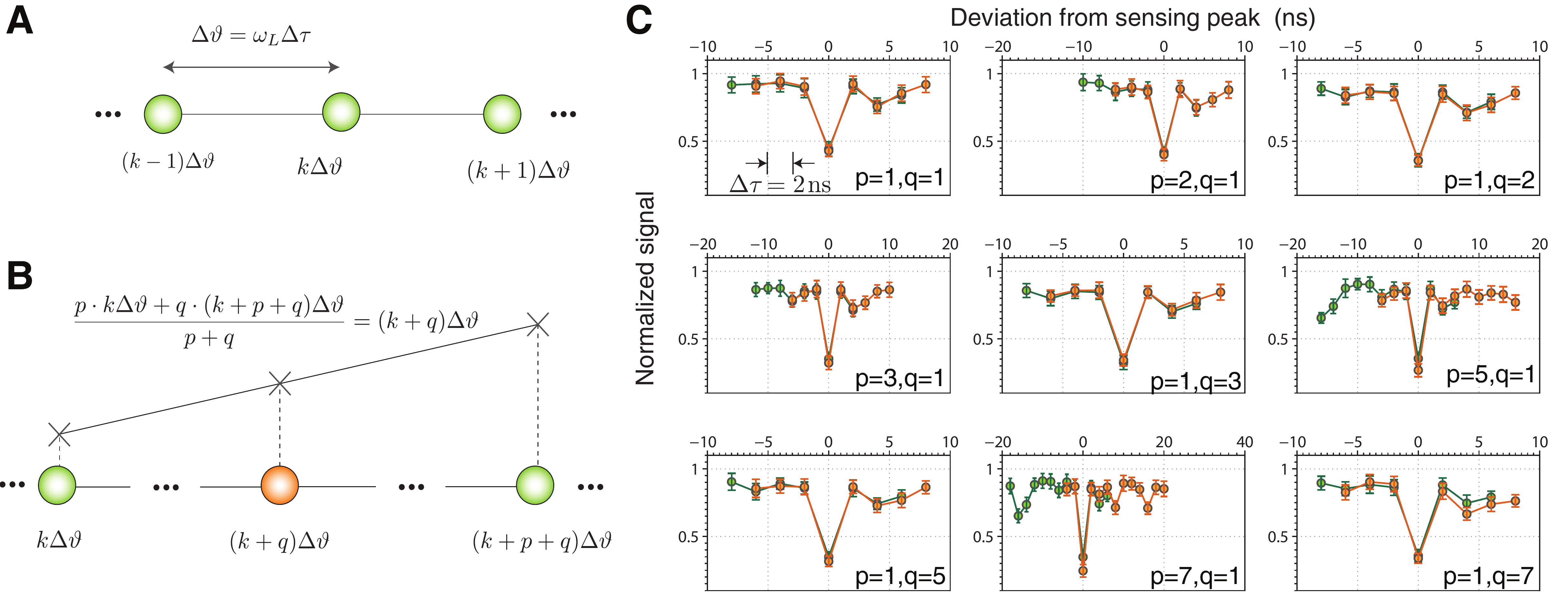}}
  \caption{\textbf{Evaluating quantum interpolation via explicit subsampling.}  In these experiments, we deliberately undersample the signal from a $\Ns$ spin using a timing resolution  that is larger than the intrinsic $\xD\qt=2$ns set by hardware. The experiments are schematically described in \textbf{(A-B)}. In \textbf{(C)}, we construct for different values of $(p,q)$ (the legend shows $q/(p+q)$) in an XY8-6 sequence,  the approximate supersampled point $\lsb U_{\ket{0}}^p(k\xD\xt+\xd)U_{\ket{0}}^q((k+p+q)\xD\xt+\xd)\rsb^{N/(p+q)}$
(orange circles) and compare it against the ideal one $U_{\ket{0}}^{N}((k+q)\xD\xt+\xd)$ (green circles). Both these propagators can be separately and individually constructed experimentally, allowing us to characterize the supersampling error with no free parameters (i.e. model independent). In these experiments $k$ is swept, and the sensing peak corresponds to the closest value of $k$ such that $k\xD\xt \app \pi$. If the construction has high fidelity error than the two lines should be identical and shifted, and hence should overlap in the panels. The results demonstrate that the construction error is low for most values of $(p,q)$, and this can be further improved by means of an optimal interpolation construction (see \zsr{optimal}).  }
\zfl{error1}
\end{figure*}
\subsection{Quantum interpolation for supersampling the sensing signal} 
\label{sec:supersampling}

The key power of the interferometric CPMG spin sensing protocol is that as the number of cycles $N$ increases the signal strength increases $\propto N^2$ and the linewidth falls as $w\propto 1/N$, yielding the double advantages of higher sensitivity and higher resolution for increasing number of pulses.  To achieve the goal of single protein structure reconstruction, we need to apply the sensing protocol in its optimal conditions, that is, at large $N$ and high field (high frequency $\xo_L$). Indeed, at high field, one also gains additionally in sensitivity and resolution because of an increase in statistical polarization of the nuclear spins being sensed~\cite{Degen09}, and the fact that parameters of interest like chemical shifts scale with magnetic field, thereby allowing an effective gain in sensing resolution.

Quantum interpolation overcomes hardware finite-timing resolution limits (see \ztr{table}) to dramatically gain in both sensitivity and resolution.

\subsubsection{Theory of quantum interpolation}

To make things concrete, consider that the nuclear signal obtained via the NV center is interferometrically obtained by sweeping the delay $2\qt$ between pulses, and
$S=\fr{1}{2} \R{Tr} \lb \mU_{\ket{0}}^{\dg}(\qt\xo_L)\mU_{\ket{-1}}(\qt\xo_L)\rb $
, where
\beq
\mU_{\ket{0},\ket{-1}}(\xt)=\mR(\xt/2, \bn_{0,1})\mR(\xt, \bn_{1,0})\mR(\xt/2,\bn_{0,1})\:,
\eeq
refer to nuclear rotation operators conditioned on the state of the NV center, with the definition $
\mR(\xt, \, \bn_j)=e^{-i \xt {\vec \sigma \cdot \bn_j }/{2}}$. The signal peak arises when the flip angle $
\xt=2\qt\xo_L\app\pi$, however due to finite timing resolution, one can only sample $\xt$ in steps of $\xD\xt=\xo_L\xD\qt$. Our technique allows one to effectively mitigate this problem through the \I{supersampling} of points, leading to a far finer grid than the $\xD\qt$ resolution barrier. To develop this notion mathematically, we denote by $U_0$ and $U_1$ two hardware defined nuclear unitaries in the $\ket{0}$ manifold of the NV center,
\beq
U_0 := \mU_{\ket{0}}(\pi + \xd_0 - \xD\xt/2)\: ; \: U_1 := \mU_{\ket{0}}(\pi + \xd_0 + \xD\xt/2)
\zl{U0U1}
\eeq
Here $\xD\xt$ refers to the effective sampling interval in angle, $\xD\qt = \xD\xt/\xo_L$, and $\xd_0$ is the parameter that describes how far away we are from the sensing peak -- in essence $\pi + \xd_0 - \xD\xt/2 = k\xD\xt$ for an integral $k$. A similar definition for \zr{U0U1} also exists in the $\ket{-1}$ subspace. Let us now assign $U_{1/2}$ the unitary arising from the product,
\beq
U_{\frac 1 2} := [U_0U_1]^{\frac 1 2} \app \mU_{\ket{0}}(\pi + \xd_0)
\zl{interpolation}
\eeq
where the subscript on $U_{1/2}$ refers to the fact we are effectively constructing a unitary, good to first order in $\xD\xt$, that lies exactly ``in-between'' the two hardware defined unitaries $U_0$ and $U_1$, and crucially which leads to the same signal as the ideal unitary $\mU_{\ket{0}}(\pi + \xd_0)$ to second order in $\xD\xt$. The high fidelity of the approximation would imply that the signal obtained is a \I{faithful} representation of the signal interferometrically obtained employing $\mU_{\ket{0}}(\pi + \xd_0)$ and $\mU_{\ket{-1}}(\pi + \xd_0)$, which cannot be accessed due to finite sampling resolution. This can be explicitly quantified as the requirement, to second order in $\xD\xt$,
\begin{widetext}
\beq
\Tr{\mU_{\ket{0}}^{\dg}(\pi + \xd_0 + \xD\xt/2)\:\mU_{\ket{0}}^{\dg}(\pi + \xd_0 - \xD\xt/2)\:\mU_{\ket{-1}}(\pi + \xd_0 - \xD\xt/2)\:\mU_{\ket{-1}}(\pi + \xd_0 + \xD\xt/2)} \app \Tr{\mU_{\ket{0}}^{2\dg}(\pi+\xd_0)\:\mU_{\ket{-1}}^{2}(\pi+\xd_0)}\zl{signal-error}
\eeq
\end{widetext}
 We will refer to the construction of \zr{interpolation} as \I{quantum interpolation} -- we have essentially interpolated the interval $\xD\qt$ by employing a composite construction of hardware accessible unitaries at the endpoints of this interval. In \zfr{trajectory}, we develop a geometric interpretation of quantum interpolation and graphically demonstrate how the error in the propagators grow with $\xD\xt$.

Generalizing this further, while hardware limits us only to sample $k\xD\xt$ and $(k+1)\xD\xt$, quantum interpolation allows us to \I{linearly} interpolate the interval $\xD\xt$ to effective {supersample} points $\lb k + \fr{q}{p+q}\rb\xD\xt$ (\zfr{supersample_panel}). Given $2N$ $\pi$-pulses in the spin sensing sequence, this can expressed as,
\bea
U_{\fr{q}{p+q}} &:=&\lsb U^p_0(k\xD\xt + \xd_0) U^q_0((k+1)\xD\xt+ \xd_0)\rsb^{\fr{N}{p+q}}\non\\
&\app& U_0^{N}\lsb\lb k + \fr{q}{p+q}\rb\xD\xt + \xd_0\rsb
\zl{supersampling-main}
\eea
Importantly, as $N$ increases, the number of points $q/(p+q)$ that can be supersampled ideally scales $\propto N$. This is remarkable because although the linewidth decreases $\propto 1/N$, the interpolated resolution scales $\propto N$, allowing one to completely mitigate the deleterious effects of timing resolution $\xD\qt$. The sensing resolution is now determined only the number of pulses that can be reliably applied, and the NV coherence time $T_2$, and experimental gains in resolution approaching three orders of magnitude are now achievable. However, for this to be true, it is crucial that the approximation infidelity in \zr{supersampling-main} is minimized. This infidelity limits the number of supersamples one can reliably construct. In the following section, we motivate a method to quantify this infidelity, and later develop an optimal construction of $U_{\fr{q}{p+q}}$.

One can draw insight about the need for an optimal construction by performing some simple experiments: the approximation becomes worse when $p\app q\app N/2$ for the same reason that the construction in \zr{supersampling-main} has an error that grows with $N$ and $\xd$. In order to study this in detail, we characterize this experimentally in \zfr{error1} via deliberate undersampling. We construct the propagators,
\beq
\lsb U_0^p(k\xD\xt+\xd)U_0^q((k+p+q)\xD\xt+\xd)\rsb^{\fr{N}{p+q}}\app U_0^{N}((k+q)\xD\xt+\xd)
\zl{error-construction}
\eeq
Both the left and right hand sides of the equation can be independently constructed  experimentally. If the approximation fidelity in \zr{error-construction} is good, the signals should be identical in both cases. The experimental results show a very good overlap, demonstrating that the construction error is low for most values of $(p,q)$. However, the error is found to slightly increase away from the sensing peak. This is an artifact of the construction of \zr{error-construction} being non-optimal. We shall show below that to minimize the error one needs to obtain a construction is one that minimizes the effective \I{distance} at which the effective error compensation occurs.

\begin{figure}[htb]
  \centering
	{\includegraphics[width=0.3\textwidth]{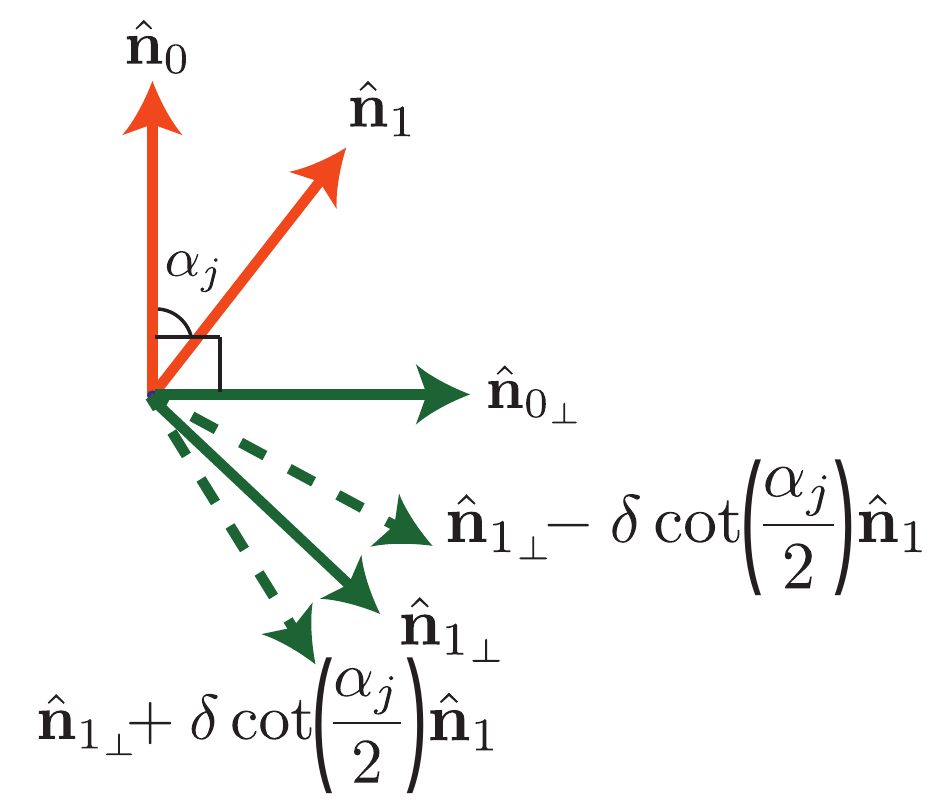}}
  \caption{\textbf{Compensation mechanism in quantum interpolation of $U_{1/2}$.} Compensation mechanism in the construction of the quantum interpolated half-way point $U_{1/2}$. We denote the two effective axes corresponding to the operators $\mU_{\ket{0}}$ and $\mU_{\ket{-1}}$ in \zr{prop}. Here the red arrows refer to the original nuclear axes $\bn_0$ and $\bn_1$ conditioned on the state of the NV, separated by the tilt angle $\xa_j$. As a result of the CPMG sequence, these axes respectively effectively become $\bn_{0_{\pp}}$ and $\bn_{1_{\pp}}$ (green arrows). To describe the linewidth of the sensing signal one notices that the propagator $\mU_{\ket{0}}(\pi\pm\xd)$ (\zr{pre-comp}) are effectively described by the vectors $\bn_{1_{\pp}} \mp\xd\cot\lb\fr{\xa_j}{2}\rb\bn_1$ (dashed green arrow), and the product  $[\mU_{\ket{0}}(\pi+\xd)\mU_{\ket{0}}(\pi-\xd)]$ points in the direction $\bn_{1_{\pp}}$ to second order in $\xd$, forming the basis of quantum interpolation.}
\zfl{compensated}
\end{figure}
	\begin{figure}[htb]
  \centering
  {\includegraphics[width=0.4\textwidth]{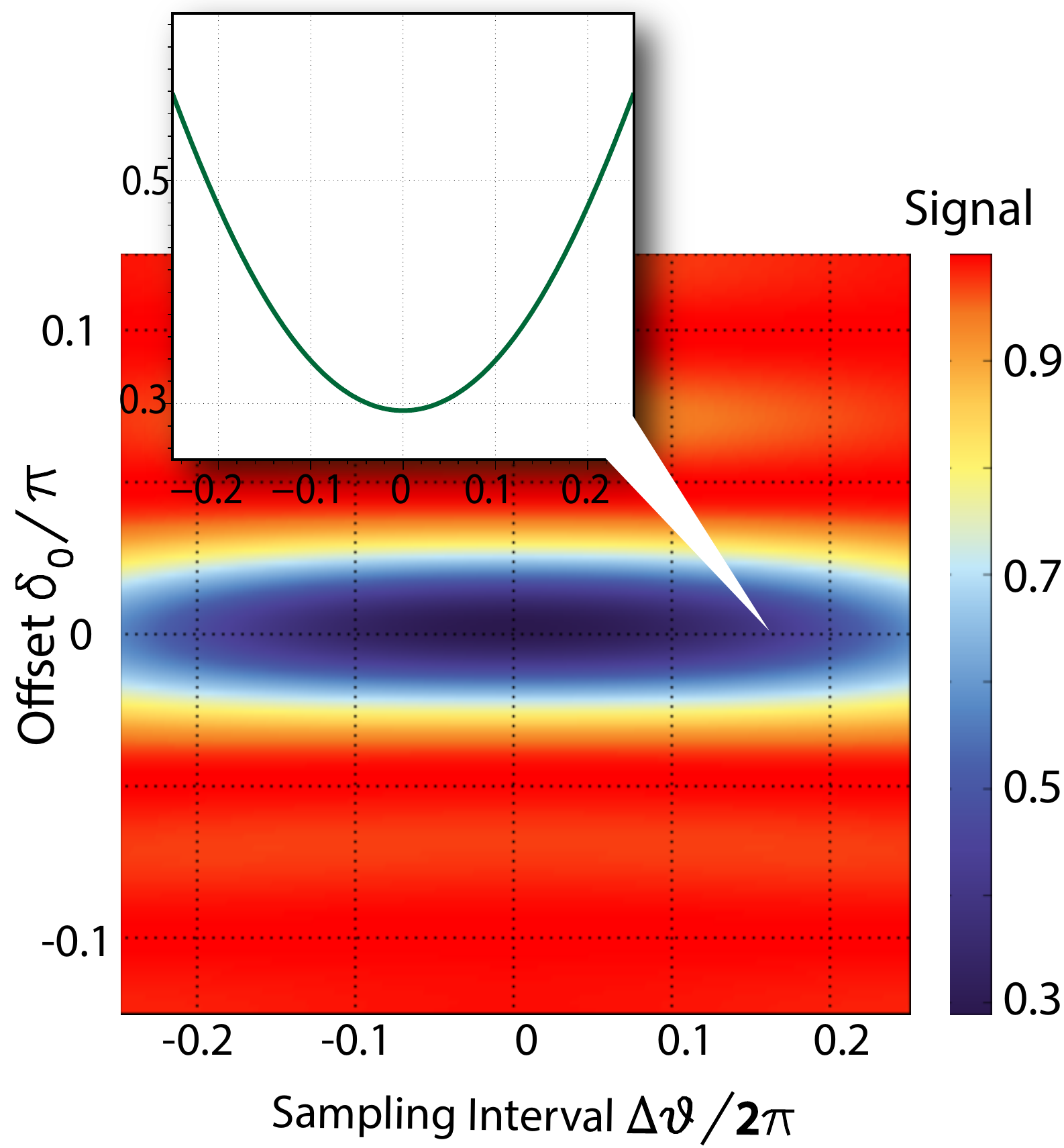}}
  \caption{\textbf{Robustness of quantum interpolation.} In this panel, we numerically study  the quantum interpolation construction for different values of $\xd_0$. More specifically, here we consider the signal obtained using the quantum interpolated unitary $U_{1/2}=[U_0U_1]^{1/2} \app \mU_{\ket{0}}(\pi + \xd_0)$, for different values of $\xd_0$. The signal is calculated with $\xa_j=0.1$ rad following \zr{signal} and is represented in color. The graph indicates, as is our expectation from \zr{signal-error}, that the interpolated unitary faithfully produces the same signal as the target unitary $\mU_{\ket{0}}(\pi + \xd_0)$ independent of the choice of $\xd_0$. Moreover, the deviation in the signal goes second order in $\xD\xt$ (inset).}
\zfl{robustness}
\end{figure}


\begin{table*}
\centering
\begin{tabular}{|c|c|c|c|c|>{\columncolor{mustard!20}}c|>{\columncolor{mustard!20}}c|}
\hline
\rowcolor{maroon!15}
\mc{1}{Instrument}&\mc{1}{Manufacturer} & \mc{1}{Timing resolution $\Delta \tau$} & \mc{1}{Jitter (Random RMS)} & \mc{1}{Cost}  & \mc{1}{$\xD\xt$ for $\Hs$ at 0.5T} & \mc{1}{Q-value for $\Hs$ at 0.5T}\\
\hline
AWG70001A\cite{AWG70001A} & Tektronix & 20 ps &  250 fs  &  \$100,000  &  $\pi/7378$ &  1174.4 \\
\hline
 AWG5002C\cite{AWG5002C}& Tektronix & 1.76 ns &  5.0 ps  &  \$32,300  &  $\pi/83$ &  13.3\\
\hline
WX1284C\cite{Tabor} & Tabor & 1 ns &  2.0 ps  & \$30,000 &  $\pi/147$ &  23.5 \\
\hline
PulseBlaster ESR-PRO \cite{Pulse} & SpinCore & 2.0 ns &  100 fs  & \$5,000 &  $\pi/73$&  11.7 \\
\hline
\end{tabular}
\caption{\textbf{Representative examples of timing instruments} commonly used for spin sensing experiments, and their respective timing resolution $\xD\qt$. There is a steep increase of instrument price with improved resolution. For the special case of a single $\Hs$ nuclear spin that is 2nm away from the NV center, and at a field of 0.5T, we specify also the corresponding values of $\xD\xt$ and the bare Q-value. Note that the Q-value is rather poor for instruments with a few nanosecond $\xD\qt$, while for sensing structural features such as chemical shifts one requires a Q-value approaching atleast $10^5$ (see \ztr{table2}). Quantum interpolation based supersampling can provide significant boosts in Q over these bare values. In the table, we also specify the RMS timing jitter of these instruments that ultimately might limit the achievable resolution via quantum interpolation. }
\ztl{table}
\end{table*}

\subsubsection{Analytical results for quantum interpolated $U_{1/2}$}
\label{sec:Uhalf}
Let us now evaluate fully analytically the quantum interpolation construction of the half-way-sample $U_{1/2}$, consisting of $2N$ pulses of the form,
\beq
U_{1/2}=\lsb \mU_{\ket{0}}(\pi + \xD\xt/2)\mU_{\ket{0}}(\pi - \xD\xt/2)\rsb^{N/2}.
\zl{compensated}
\eeq
This can be effectively translated to a pulse sequence following Fig. 1\textbf{(A)} of the main paper, consisting of pulses that are unequally spaced. To illustrate the mechanism of the interpolation in \zr{compensated}, consider that we had determined that the origin of the finite linewidth of the spin sensing sequence is the fact that away from the signal peak $(\xd=0)$ there is effectively destructive interference in the flip angle $\xa_j$ into $\xa_j'$ (see \zr{alpha-prime}) leading to a loss of signal contrast. Now if the effective flip angle $\xa'_j$ can be made \I{independent} of $\xd$, then the signal linewidth is proportional to a constant to second order. This is precisely what occurs through the quantum interpolation construction in \zr{compensated}.

To describe this in detail, for simplicity, once again one can use a first order expansion to discern the physics of the problem. Evaluating the effective propagators, for effectively $N/2$ cycles (keeping the number of pulses the same as in \zr{pre-comp}), we obtain

\begin{widetext}
\bea
\lsb\mU_{\ket{0}}(\pi + \xD\xt/2)\mU_{\ket{0}}(\pi - \xD\xt/2)\rsb^{N/2} &\app&  \id\cos N\xa_j + i\fr{\sin N\xa_j}{\sin 2\xa_j}\:\boldsymbol\xs\cdot \lsb -\bn_{1_{\pp}}(2\cos\xa'_j\:\sin\xa_j) - \xD\xt\sin\xa_j\:(1+\cos\xa_j)(\bn_{1_\pp}\zt\bn_{1})\rsb\non\\
\lsb\mU_{\ket{-1}}(\pi + \xD\xt/2)\mU_{\ket{-1}}(\pi - \xD\xt/2)\rsb^{N/2} &\app&  \id\cos N\xa_j + i\fr{\sin N\xa_j}{\sin 2\xa_j}\:\boldsymbol\xs\cdot \lsb \bn_{0_{\pp}}(2\cos\xa'_j\:\sin\xa_j) + \xD\xt\sin\xa_j\:(1+\cos\xa_j)(\bn_{0_\pp}\zt\bn_{0})\rsb\:
\label{eqn:robust-cons}
\eea

\end{widetext}
Remarkably the dependence of $\xa'_j$ on the $\xD\xt$ and the destructive interference in \zr{alpha-prime} is now removed (compare with \zr{robust-cons} with \zr{pre-comp}). The sensing signal as a function of $\xD\xt$ is now
\bea
S &=& \cos^2N\xa_j - \sin^2N\xa_j\cos\xa_j\non\\
&-& \fr{\sin^2N\xa_j}{\sin^22\xa_j}\:\lsb\xD\xt^2\sin^2\xa_j\:(1+\cos\xa_j)^2\:(1-\cos\xa_j)\rsb.
\zl{second-order}
\eea
The first line is exactly the signal magnitude obtained at the signal peak -- but now the width is set weakly by the second line that goes as $\xD\xt^2$. 
This quantum interpolation compensation mechanism has also a simple geometric interpretation (\zfr{compensated}) -- similar to a spin echo~\cite{Hahn50}, the linear dependence on $\xD\xt$ leading to the destructive interference in \zr{alpha-prime} is removed by employing another vector with the opposite sign, giving an effective propagator that is independent of $\xD\xt$. 

We note that while in the above analysis we considered the case of the offset $\xd_0=0$ in \zr{U0U1}, we can also numerically evaluate that the quantum interpolation construction for $U_{1/2}$ is robust to different values of offset $\xd_0$. This is shown in \zfr{robustness}, where the shading represents the signal obtained as a function of $\xD\xt$ for different $\xd_0$. It is evident that for any slice in the $\xd_0$ dimension, the signal falls off quadratically in $\xD\xt$, a reflection of the fact that to first order the quantum interpolation compensation mechanism (\zfr{compensated}) is still robust.

\subsubsection{Evaluating the fidelity of quantum interpolation}
As a clarifying calculation, let us evaluate how close the unitary out of quantum interpolation is close to the ideal one. Using a trace norm measure, we show that this is approximately second order in $\xD\xt$. The ideal propagator is $U_{id} = \mU^2_{\ket{0}}(\pi)= \exp(-i2\xa_j\xs\cdot\bn_{1\pp})$. Comparing with the quantum interpolated expression, we  have that the trace norm,
\bea
F&=&\Tr{\mU_{\ket{0}}(\pi + \xD\xt/2)\mU_{\ket{0}}(\pi - \xD\xt/2)U_{id}^{\dg}} \non\\
&=&\cos2\xa_j\:[\sin^2(\xD\xt/2) + \cos^2(\xD\xt/2)\cos2\xa_j]\non\\
&+& 2\sin2\xa_j\:\sin\xa_j\:\cos\xa'_j\cos(\xD\xt/2)\non\\
&\app& 1 + {\cal O}(\xD\xt^2)
\zl{fidelity-second-order}
\eea
Note that except the $\cos\xa'_j$ term, all the other terms are second order or more in $\xD\xt$. The $\cos\xa_j'$ term too is weighted by $\sin^2\xa_j$, and for most practical cases of spin sensing where $\xa_j$ is small, this term has a negligible contribution. Hence to a very good approximation, the quantum interpolation expression is good to first order in $\xD\xt$. A graphical comparison of the unitaries obtained via quantum interpolation to the ideal one is shown in \zfr{trajectory}.

\subsubsection{Comparison with Baker-Campbell-Hausdorff result}
Let us now demonstrate that the effectiveness of the quantum interpolation construction cannot be seen as a simple manifestation of the zeroth order of the Baker-Campbell-Hausdorff (BCH) expansion~\cite{BCH,Magnus54}. The zeroth order BCH expression does not care for cross terms or commutators between the two unitaries, and can be written down as, $\lsb\mU_{\ket{0}}(\pi + \xD\xt/2)\mU_{\ket{0}}(\pi - \xD\xt/2)\rsb \app U_{\R{BCH}}$, with,

\bea
U_{\R{BCH}}&=& \exp\lsb -i\xa'_j\boldsymbol\xs\cdot(\bn_{1+} + \bn_{1-})\rsb\non\\
&=& \exp\lsb i2\xa'_j\boldsymbol\xs\cdot \fr{-\bn_{1\pp}\sin\xa_j\:\cos(\xD\xt/2)}{\sin\xa'_j}\rsb
\zl{trotter}
\eea
where $\bn_{1\pm}$ are the \I{exact} effective vectors in the expressions for unitaries away from the sensing peak (\zr{pre-comp}),
\begin{align}
\bn_{1+} &=& \fr{1}{\sin\xa_j'}\lsb \bn_{1\pp}\sin\xa_j - \bn_{0}\sin(\xD\xt/2)\:(1+\cos\xa_j)\rsb\cos(\xD\xt/2)\non\\
\bn_{0+} &=& \fr{1}{\sin\xa_j'}\lsb \bn_{0\pp}\sin\xa_j - \bn_{1}\sin(\xD\xt/2)\:(1+\cos\xa_j)\rsb\cos(\xD\xt/2)
\zl{exact-prop2}
\end{align}
For $N/2$ cycles of the CPMG experiment, this has the form,
\beq
U^{N/2}_{\R{BCH}} \app \exp\lsb iN\xa_j'\cos(\xD\xt/2)\fr{\sin\xa_j}{\sin\xa_j'} (\boldsymbol\xs\cdot -\bn_{1_{\pp}})\rsb
\zl{trotter1}
\eeq
\zr{trotter1} immediately reveals that the quantum interpolation compensation effect in \zr{robust-cons} cannot be captured by a simple BCH analysis. This is because $\xD\xt$ cannot be seen as a perturbative parameter in the expressions, and in general the BCH expansion does not converge~\cite{Haeberlen76}. More intuitively, the flip angle compensation in  \zr{alpha-prime} that was crucial to remove the dependence of $\xd$ to bring back the bare flip angle of $\xa_j$ in \zr{robust-cons} is no longer present. Instead the flip angle is now $\xa_j'\cos(\xD\xt/2)\fr{\sin\xa_j}{\sin\xa_j'}$, which only approaches the right expression when $\xa_j'$ is small. Note however that we have made no assumptions in our analysis about $\xa_j$ being small, and hence the simple zeroth order BCH analysis leads to a larger error than a more complete analysis that also includes the effect of commutators or cross terms.

\subsubsection{Survey of hardware and comparison with supersampling}
\label{sec:survey}
\ztr{table} summarizes a list of hardware used for spin sensing experiments, and their respective timing resolutions. In our experiments in \zsr{expt} we used the SpinCore PulseBlaster~\cite{Pulse} and Tabor Arbitrary Waveform Generator~\cite{Tabor} with timing resolutions of 2ns and 1ns respectively. Using the latter instrument, through quantum interpolation based supersampling we were able to experimentally demonstrate a resolution of 8.9ps, a boost by a factor of 112 (Fig. 3 of the main paper).   Note that random timing jitter of these instruments sets the ultimate achievable resolution through supersampling.

\begin{figure}[bht]
  \centering
  {\includegraphics[width=0.375\textwidth]{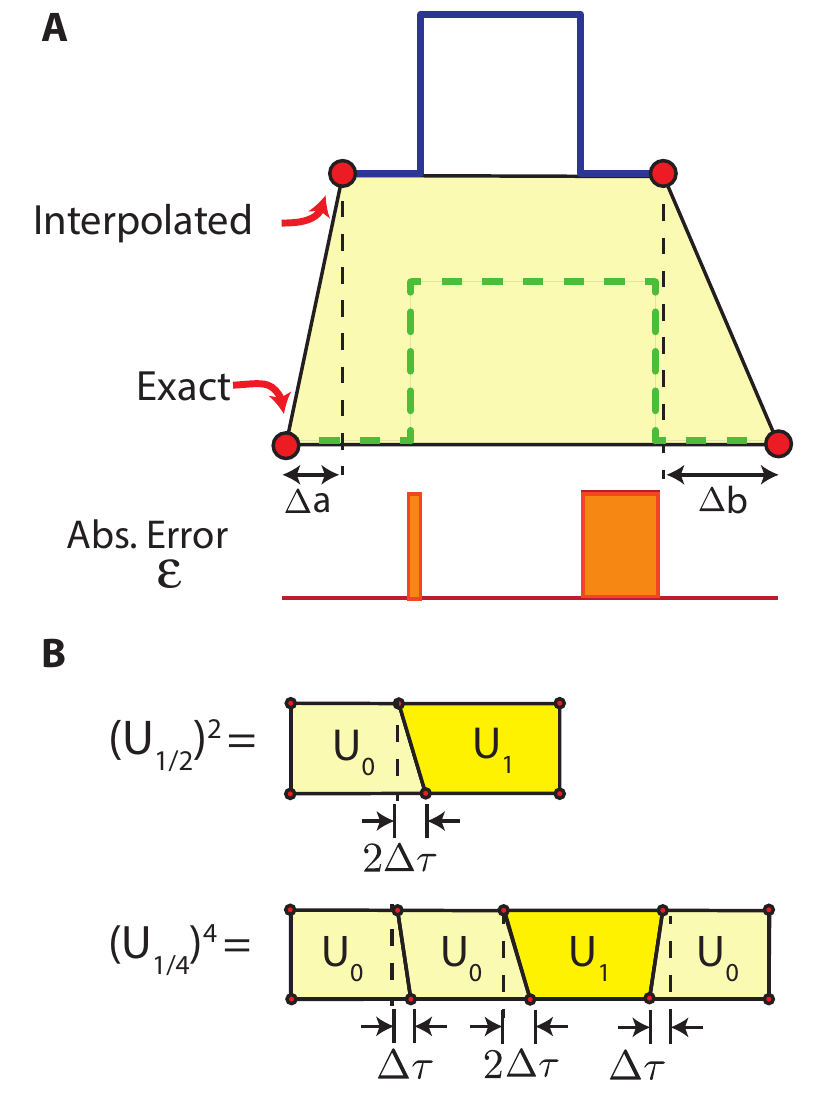}}
  \caption{\textbf{Evaluating construction error via the filter formalism.} \textbf{(A)} Here the upper rail of the trapezium represents the interpolated construction (approximate), and the lower rail represents the target construction (ideal) (see also \zfr{circle}). The  circles denote the \I{total} period of the corresponding interpolated (blue line) and exact time domain filter functions (green dashed lines), corresponding to Fig. 1\textbf{(C)} of the main paper. $\xD a$ and $\xD b$ denote the deviations of the interpolated filter from the exact one, and the net error is then $\xe = \left| -\fr{3}{4}\xD a + \fr{1}{4}\xD b\right| + \left| -\fr{1}{4}\xD a + \fr{3}{4}\xD b\right|$, which is minimized by the optimal supersampling construction. \textbf{(B)} Optimal constructions for the quantum interpolated points one-half and one-quarter between two hardware defined samples (see also \zfr{error-matrix}).}
\zfl{trapezium}
\end{figure}

\begin{figure}[htb]
  \centering
	\includegraphics[width=0.45\textwidth]{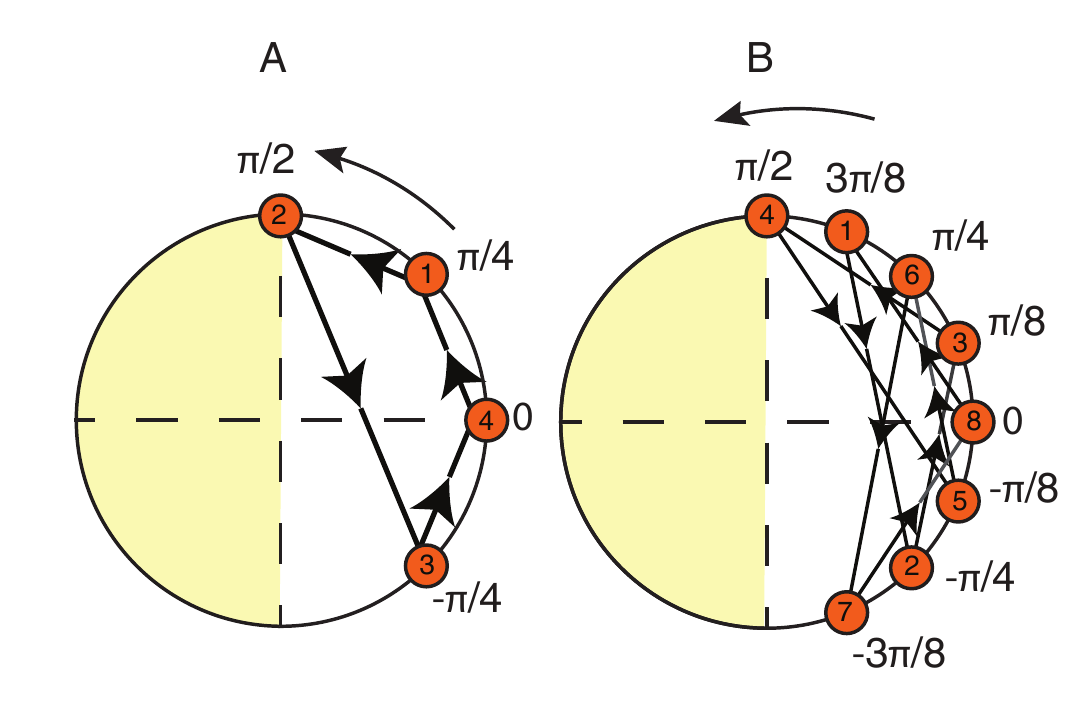}
  \caption{\textbf{Geometric representation of the optimal construction,}  for two particular sampling times \textbf{(A)} $\xD\qt/4$ and \textbf{(B)} $3\xD\qt/8$. We represent the desired sample as phasor at angle $\pi \zt \R{sample}$ on a circle (orange circles). The left half plane (LHP) of the circle (shaded) is considered forbidden. Starting with an initial loop counter $m=0$,  we propagate the algorithm by forming the phasor $m\rt m+\R{sample}$ (arrows); and so long as we don't pass into the LHP we assign to this the operator $U_0$. In the opposite case, we assign $U_1$, and reflect the phasor about the origin.  The algorithm ends when we finally return to the starting position. The numbers in the orange circles indicate the progression of the algorithm.  This geometric representation also allows an intuitive understanding of why the error of all samples is almost the same (see \zfr{optimal-construction-graph}).}
\zfl{circle}
\end{figure}			


\begin{figure}[htbp]
\begin{algorithm}[H]
\caption{: Construction of the optimal interpolation sequence}
\begin{algorithmic}
\Procedure{OptimalConstruction}{}
\State Set loop iteration counter $m \gets 0$
\State Optimal sequence string $U\gets \R{null}$
\BState \emph{\textbf {loop}}:
\State \I{Propagate} $m \gets m+ \text{sample}$
\If {$|m| \leq 1/2$} Append $U_0$ operator to sequence
\Else
\State Append $U_1$ operator to sequence
\State \I{Reflect} $m\gets m-1$
\EndIf
\While {$m\neq 0$} \State{\textbf{goto} \emph{\textbf {loop}}} \EndWhile
\BState \textbf{end procedure}
\EndProcedure
\end{algorithmic}
\end{algorithm}
\caption{\textbf{Algorithm for optimal quantum interpolation construction.} The algorithm produces the optimal sequence of $U_0$ and $U_1$ operators to interpolate the desired sampling point. In the algorithm, ``sample'' stands for a fraction between 0 and 1 corresponding to the desired supersample time.  See  Sec.~\ref{sec:code} for an explicit MATLAB implementation of this algorithm.  }
\zfl{algorithm}
\end{figure}

\begin{figure*}[htbp]
  \centering
	\includegraphics[width=0.95\textwidth]{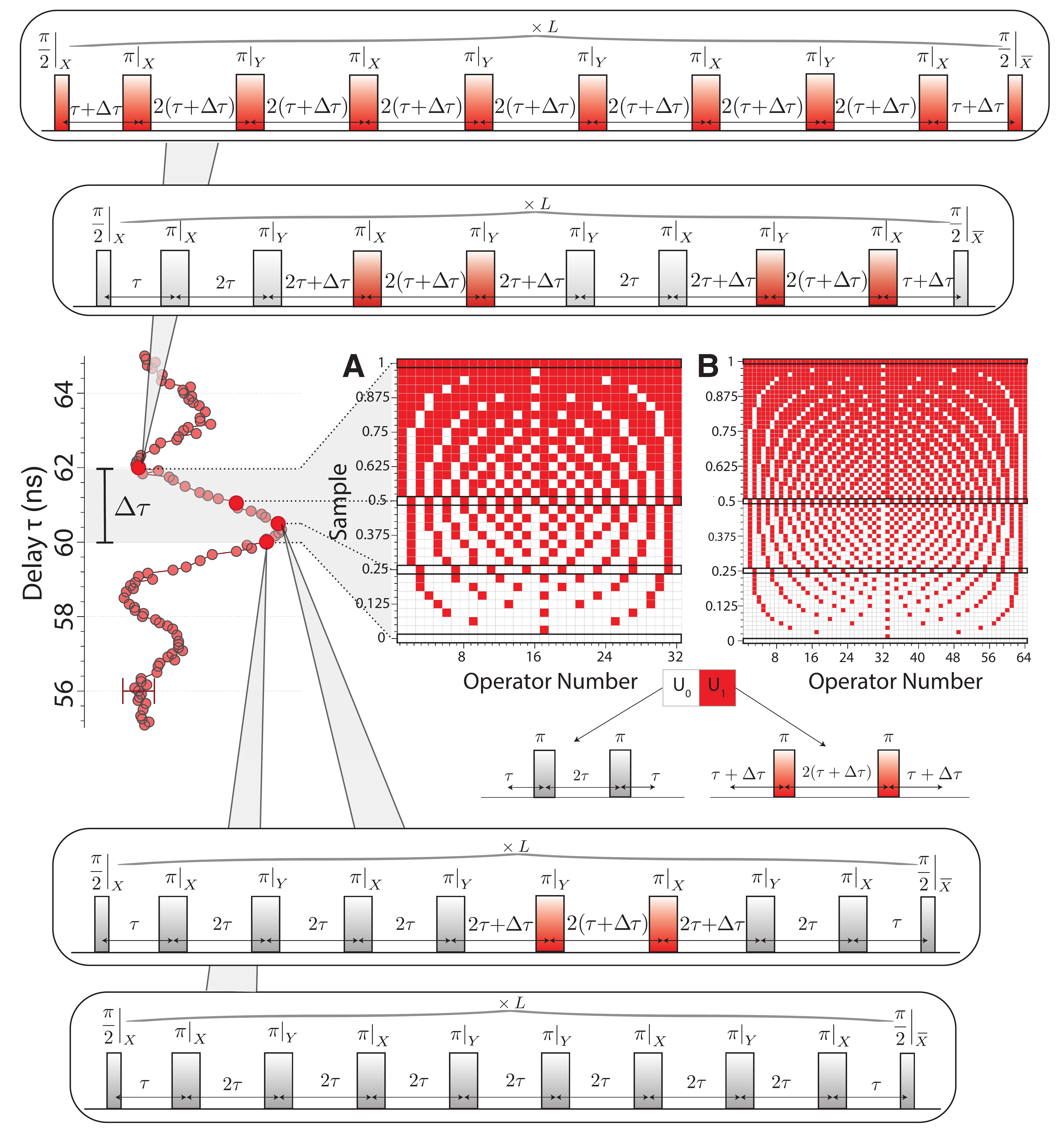}
  \caption{\textbf{Pulse sequence construction for optimal quantum interpolation.} Shown is the optimal quantum interpolation supersampling construction for different samples for \textbf{(A)} N=4 and \textbf{(B)} N=8 cycles of an XY8 sensing sequence. Here ``sample'' refers to point to be interpolated between two hardware limited intervals as the fraction of the timing resolution $\xD\qt$. For reference example experimental data is shown (red circles, corresponding to Fig. 1\textbf{(B)} of the main paper). Here $\xD\qt=2$ns, and in each such interval one could effective supersample proportional to the number of pulses applied. To experimentally construct the optimal interpolated sequence for a sample corresponding a particular row of the matrix, one applies the sequence of operators $U_0$ (white) or $U_1$ (red) from left to right. The boxed inset panels
	denote the pulse sequences corresponding to four example samples -- $\{0,1/4,1/2,1\}$.}
\zfl{error-matrix}
\end{figure*}
\begin{figure}[htbp]
  \centering
		\includegraphics[width=0.5\textwidth]{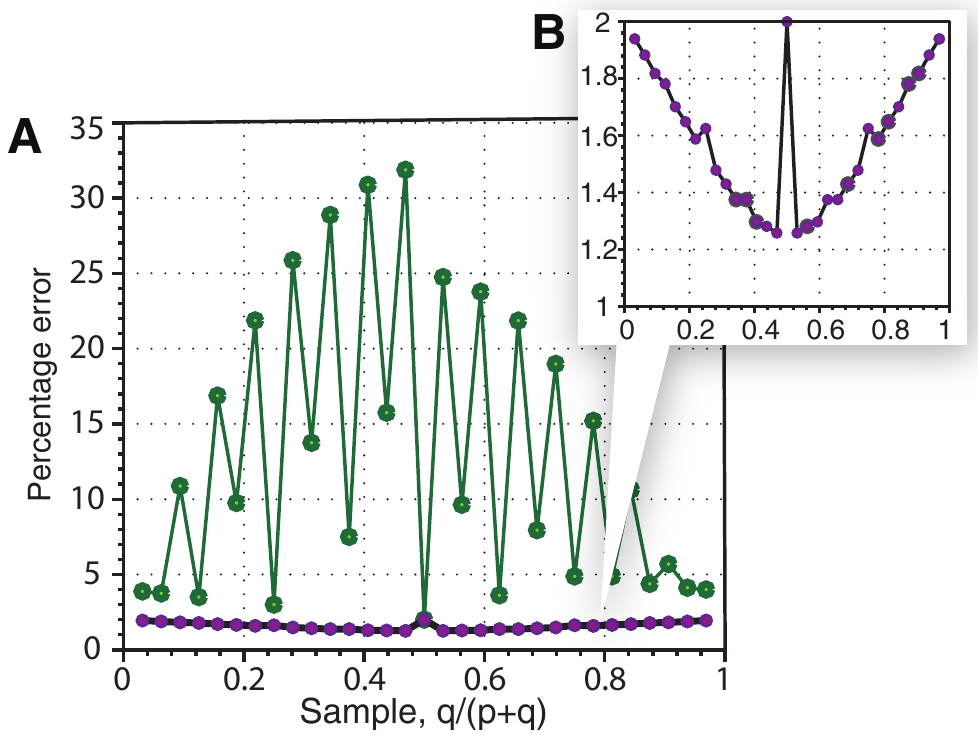}
  \caption{\textbf{Performance of optimal quantum interpolation.}  We represent the percentage errors (specified in units of $\xD\qt/4\qt$) of the construction of a supersample $\lsb q/(p+q)\rsb \xD\qt$ via quantum interpolation using two hardware defined unitaries $U_0$ and $U_1$, and an time resolution of $\xD\qt$. Here we consider $N=16$ cycles of the CPMG sensing sequence, giving a total of 32 possible supersamples. The error is calculated from the effective area under the time domain error function $\xe$ following \zfr{trapezium}. \textbf{(A)} The green points denote the naive construction $U_0^pU_1^q$, where the errors accumulate very quickly. The purple line and points instead denote the case of the optimal construction following \zfr{circle}, where the error of all samples is less than the half-way-sample (zoomed in the inset \textbf{(B)}). Hence the optimal construction can reliably produce different supersamples with low effective error (see also \zfr{optimal-error}).}
\zfl{optimal-construction-graph}
\end{figure}

\begin{figure}[htbp]
  \centering
  {\includegraphics[width=0.5\textwidth]{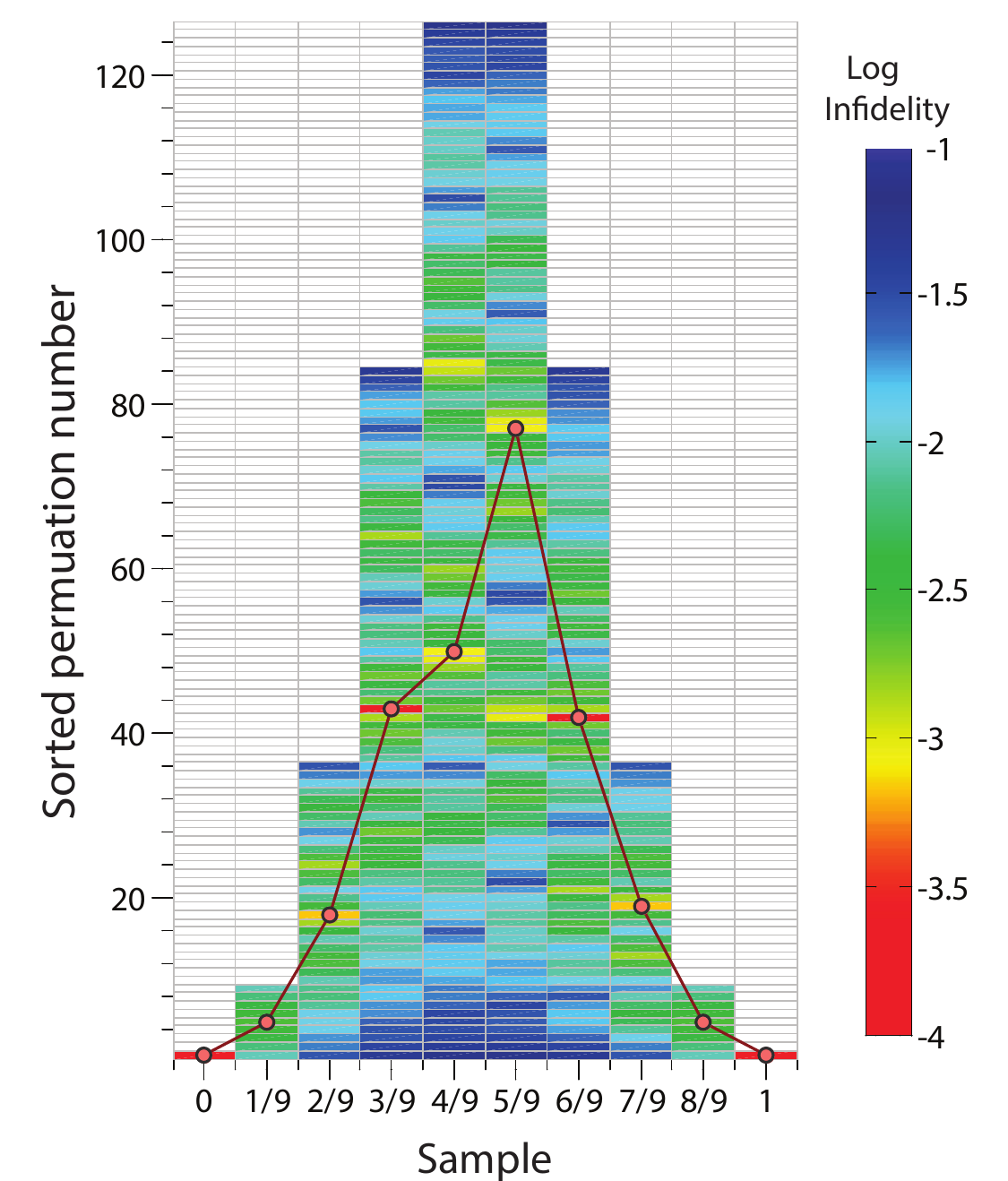}}
  \caption{\textbf{Evaluating optimality of quantum interpolation constructions.} Figure evaluates the optimal quantum interpolation construction for $N=9$ cycles of the CPMG sequence. For every sample, of the form $p\xD\qt/N$, for integral $p$, we evaluate the fidelity of the interpolated unitary obtained from hardware defined unitaries $U_0$ and $U_1$ to the target unitary, for \I{every} possible permutation of $U_0$ and $U_1$. For clarity, for instance, for the sample 1/9, there are 8 permutations of the sequence $[U_0U_1^8]$, and we study the infidelity of each of these permutations (second bar). Here we evaluated the mean infidelity over the range $\xd_0\in [-2\xD\xt,2\xD\xt]$, for $\xD\xt=\pi/20$ and $\xa_j=0.1$ rad. The permutations are sorted down to up by Hamming weight, i.e. in increasing decimal order of their sequence strings. The colors represent the log of the infidelity, and the smaller number represents that the constructed unitary is better, i.e. has lower error. The optimal construction obtained using the Algorithm in \zfr{algorithm} are shown in the by the orange circles. Numerically, we find that the construction from  \zfr{algorithm} does indeed capture the optimal possible permutation.  }
\zfl{optimal-error}
\end{figure}
\begin{figure*}[htb]
  \centering
	{\includegraphics[width=1\textwidth]{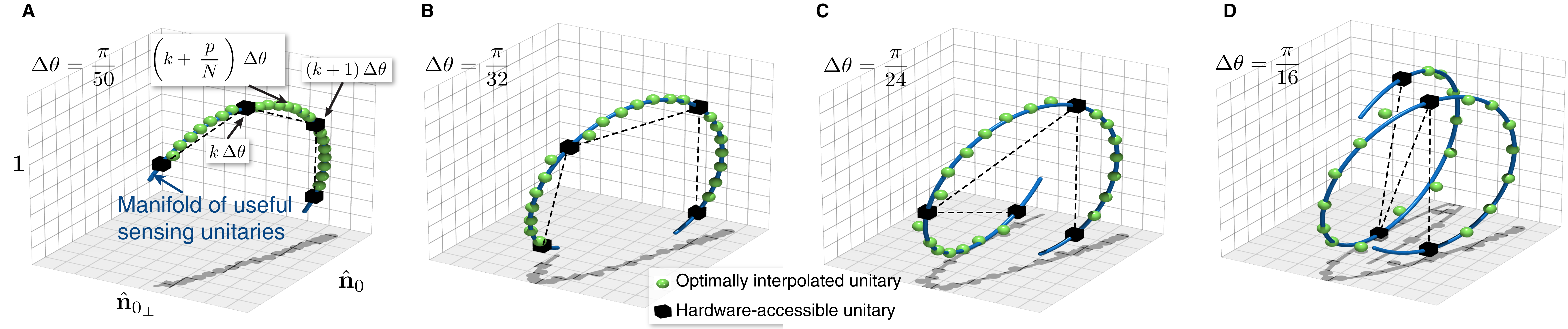}}
  \caption{\textbf{Quantum interpolation from a geometric perspective.}  The blue line indicates the ideal manifold of sensing unitaries, while the black squares represent the hardware accessible unitaries (see also Fig. 1\textbf{(A)} of the main paper).  
We represent unitaries as points in the 3D space spanned by the operators $\{\id,\bn_0,\bn_{0_{\pp}}\}$. The projection of the missing dimension, $(\bn_0\zt\bn_{0_\pp})$ is small and only causes the norm of the vector to be smaller than one.
Without quantum interpolation, the signal obtained upon sweeping $\xd_0$ corresponds to the dashed black line. The green circles are the unitaries resulting from quantum interpolation with two successive hardware accessible samples following the optimal construction, where one seeks to construct the ideal unitaries $\mU^N_{\ket{0}}(\pi-\xd_0+k\xD\xt)$, with $\xd_0\in\{-\xD\xt,0,\xD\xt,2\xD\xt\}$. Here $N=8$ leading to 8 supersamples in each $\xD\xt$ interval, and we considered $\xa_j=0.1$ rad. The different panels denote different values of $\xD\xt$. The success of quantum interpolation is evident from the fact that the two result of interpolation closely matches the target unitaries over the entire manifold, even for large  $\xD\xt$.}
\zfl{trajectory}
\end{figure*}

\subsection{Optimal supersampling construction}
\label{sec:optimal}
The  naive construction based on the approximation of \zr{supersampling-main} is not optimal, and carries the error that finally limits the number of additional supersamples that reliably constructed. Thus we develop an optimal construction for quantum interpolation to overcome this problem. The optimal construction sets the \I{order} of operators $U_0$ and $U_1$  used to interpolate a supersample $q\xD \qt/(p+q)$ with the lowest amount of error.

\subsubsection{Error in sequence construction: Semiclassical analysis}

The simplest method to characterize the error of supersampling sequences is through a semiclassical analysis using the filter formalism of dynamical decoupling~\cite{Viola98,Cywinski09,Alvarez10b,Silva14}, as it enables a simple optical analogy~\cite{Ajoy13}. Here we assume a classical noise field acting on the NV center, yielding the Hamiltonian $\mH_n = bE_z(t)S_z$. Here $E_z(t)$ is a classical noise field, assumed to be Gaussian-distributed with zero mean. 
For instance, $E_z(t)$ might approximate the spin noise for an ensemble of weakly coupled nuclear spins. 
For stationary noise, the time-correlation is $\expec{E_z(t)E_z(t+\qt)} = g(\qt)$, with the noise spectral density $S(\xo)=\fr{1}{\sq{2\pi}}\int_{\infty}^{\infty}dtg(t)e^{-i\xo t}$. 
For example, the spectral density function due to nuclear spin noise is centered at their resonance frequency, with zero linewidth if considering a single  nuclear spin. 
In the toggling frame, each $\pi$ pulse in the control sequence flips the sign of the noise Hamiltonian $\mH_n$, leading to the effective time-dependent Hamiltonian $\wt{\mH} = bf(t)E_z(t)S_z$. 
The time-domain filter function $f(t)$ switches between $\pm 1$ at each pulse. 
The decay of the coherence of the NV center is then given by the overlap integral $\chi(t) = \fr{\sq{2\pi}|b|^2}{2}\int_{-\infty}^{\infty} |F(\xo)|^2S(\xo)d\xo$, between the frequency domain filter $F(\xo)$ (the Fourier transform of $f(t)$) and the noise spectrum.

Now, given the finite timing resolution $\xD\qt$, we can only obtained two different time-domain filter functions with total time separated by  $4\xD\qt$. The aim of quantum interpolation is to obtain a filter that leads to the same signal as the effective filter that is ``in-between'' these two hardware separated filters. While the error arises from differences between the ideal and interpolated $F(\omega)$, to evaluate the how closely this construction is a faithful representation of the ideal filter, by Parseval's theorem, one just needs to determine the deviation $\xe$ of the interpolated time-domain filter from the ideal one. 

For the case of $U_{1/2}$, the relative error is  just $\xD\qt/2\qt$, which is proportional to the size of the sampling interval.
This provides a convenient starting point to determine the \I{optimal} interpolation construction for any arbitrary sampling point $\fr{q}{p+q}\xD\qt$: Essentially the optimal construction is the one that minimizes the \I{net} deviation $\xe$ of the time-domain interpolated filter from the ideal one. \zfr{trapezium} offers a simple prescription to calculate this error; the upper rail represents the filter corresponding to the quantum interpolation construction out of $U_0$ and $U_1$ operators, while the lower rail represents the ideal filter. The total length for both rails is identical -- this ensures that the filter does indeed sample the correct frequency. Comparing each filters for each successive application of operators (i.e. piece wise), one obtains trapezium shaped blocks that can be pieced together to evaluate the error of a supersampling sequence. The net error of each of these blocks has the form (see \zfr{trapezium}),
\beq
\xe = \left| -\fr{3}{4}\xD a + \fr{1}{4}\xD b\right| + \left| -\fr{1}{4}\xD a + \fr{3}{4}\xD b\right|
\zl{trap-error}
\eeq
The optimal constructions, for instance shown in the lower panels of \zfr{trapezium} for $U_{1/2}$ and $U_{1/4}$ minimize this error. Note that the error in \zr{trap-error} is maximized in the situation where $\xD a$ is negative, and $\xD b$ positive, giving the bound,
\beq
\xe \leq (\xD a + \xD b)
\zl{trap-error2}
\eeq

\begin{figure}[htb]
  \centering
			  {\includegraphics[width=0.45\textwidth]{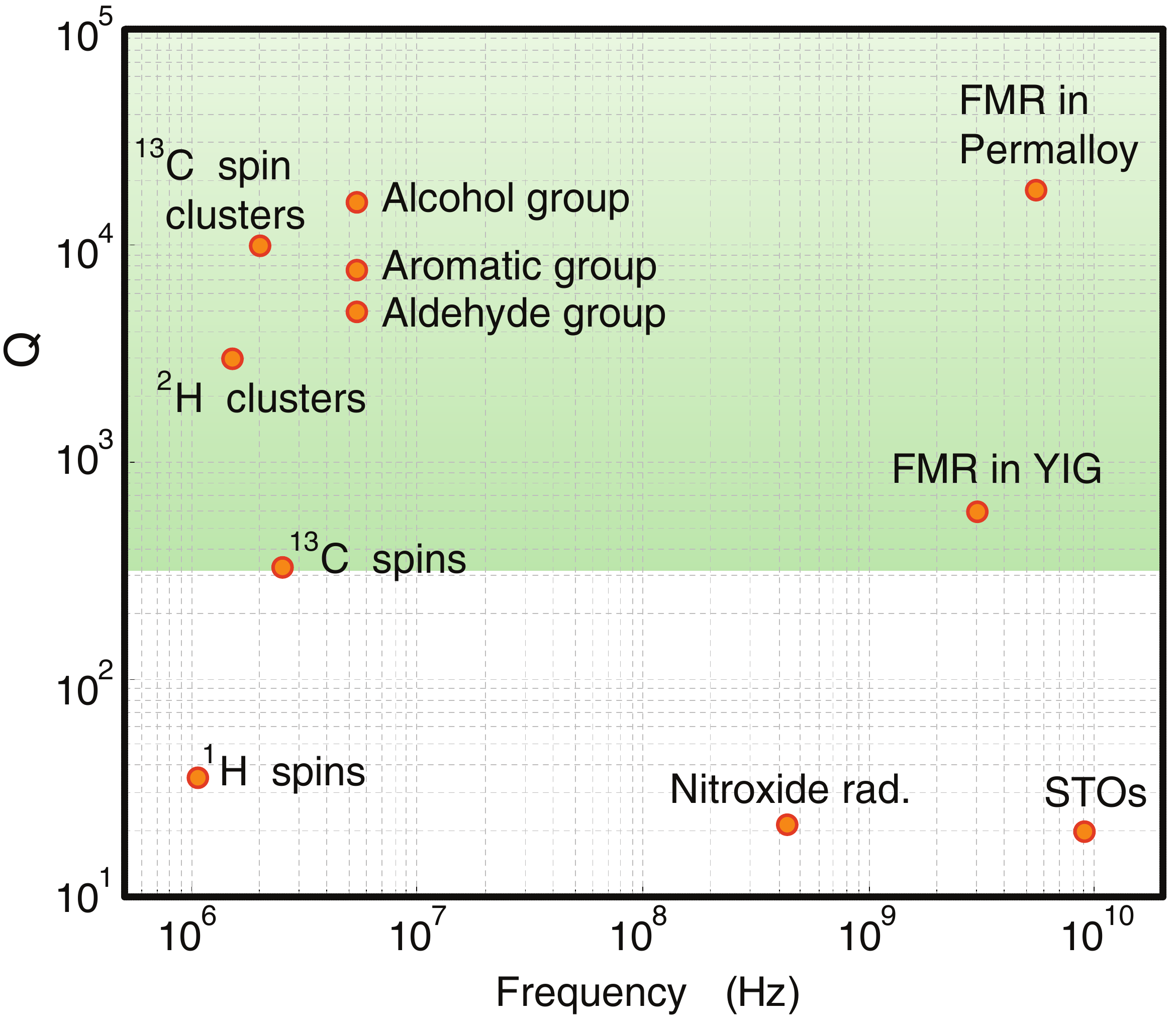}}

                          \caption{\textbf{Q-value vs $f$ for different quantum metrology experiments.} In this scatter plot we show representative examples the Q-values from different signal sources plotted against their native frequency $f$ (see \ztr{table2}). Note that in our experiments, we could obtains gains in Q by about $10^4$ via quantum interpolation, and this might open up the metrology of several signals (green shaded region) that were hitherto extremely challenging.   }
\zfl{scatter}
\end{figure}
\begin{table*}[htbp]
\centering
\begin{tabular}{|c|c|c|c|>{\columncolor{mustard!20}}c|c|>{\columncolor{mustard!20}}p{6cm}|}
\hline
\rowcolor{maroon!15}
\mc{1}{Experiment/System}&\mc{1}{Signal Source}&\mc{1}{f} & \mc{1}{$\xD f$} & \mc{1}{Q-value} & \mc{1}{Reference} & \mc{1}{Notes}\\
\hline

\multirow{5}{*}{\parbox{3cm}{Nanoscale NMR/ESR}}& $\Cs$ spins & 2.5 MHz & 7.5 kHz &  333 &  \cite{Lovchinsky16}& \parbox{6cm}{Nanoscale spin sensing experiments on a single protein of Ubiqutin.}\\
    &Nitroxide rad. & 430 MHz & 20 MHz &  21.5  &  \cite{Shi15}& \parbox{6cm}{Electron radicals measured in a single protein.}\\
  &$\Hs$ spins & 1.06 MHz & 30 kHz &  35.3  &  \cite{Staudacher15}& \parbox{6cm}{Proton spins measured in an organic molecule outside diamond.}\\
	&\parbox{3cm}{$\Cs$ spin clusters} & 2 MHz & 200 Hz &  $10^4$  &  \cite{Ajoy15}& \parbox{6cm}{Simulated linewidths requried to resolve clusters of spins in CXCR4.}\\
	&\parbox{3cm}{Nanoscale NQR of ${}^{2}$H clusters} & 1.5 MHz & 500 Hz &  3000  &  \cite{Lovchinsky16}& \parbox{6cm}{Simulated linewidths requried to resolve NQR peaks arising in deuterated phenylalanine.}\\
	\hline
	\multirow{3}{*}{$\Cs$ Chemical shift}& Aldehyde group & \multirow{2}{*}{5.35 MHz at 0.5T} & \multirow{2}{*}{-} &  $\sim$ 5000  &  \multirow{2}{*}{\cite{Ernst,chemshift}}&\parbox{6cm}{We assume that to be able to resolve chemical shifts, one requires a Q of atleast the shift value.}\\
    &  Aromatic group &  &   &  $\sim$ 7700  &  &\\
		&  Alcohol group &  &   &  $\sim$ 16000  &  &\\\hline
\multirow{3}{*}{\parbox{3cm}{Spin waves in Ferromagnets}}& FMR in Permalloy & 5.5 GHz & 0.3 MHz &  $1.8\zt 10^4$ &  \cite{van15}& \parbox{6cm}{Here we consider \I{spectrally} sensing the spin waves excitations directly via the NV.}\\
    &FMR in YIG & 3 GHz & 5 MHz &  600 &  \cite{van15,trifunovic15}& \parbox{6cm}{}\\
		&STOs & 9 GHz & 450 MHz &  20  &  \cite{Liu12s}& \parbox{6cm}{Spin torque switching in Tantalum.}\\
		\hline

\hline
\end{tabular}
\normalsize
\caption{\textbf{Examples of Q-values from different sources.} In this table we show representative examples of Q-values required to effectively sense signals from different sources, including single spins, chemical shifts, and spin wave modes in ferromagnets (see also \zfr{scatter}). We note that most measurements in the literature are of low Q-value, below 1000 (see also \ztr{table}).  Given that we can experimentally achieve substantial gains in Q-value due to quantum interpolation, several of the high Q signals are now within the regime of quantum metrology with NV centers.}
\ztl{table2}
\end{table*}

\subsubsection{Algorithm for optimal supersampling construction}
Let us now determine the supersampling construction that minimizes the error $\xe$ in \zr{trap-error} -- the deviation from the ideal filter. For a sample of the form $\R{sample} = q/(p+q)$, we obtain the optimal string of operators $U_0$ or $U_1$ with the minimization evaluated at the end of each applied operator following \zfr{trapezium}. We note that while in principle one has to minimize the deviation of the time domain filters edge to edge in \zfr{trapezium}, however it is sufficient to use a simple approach of minimizing deviations in the total periods.  A simple algorithm that achieves this has the pseduocode in \zfr{algorithm}. The optimal construction is shown in \zfr{error-matrix}, where the colors represent the operators $U_0$ or $U_1$. The panels describe the construction of different supersamples, the total number of which scales linearly with the number of pulses (shown are the examples of XY8-4 and XY8-8). 

 A geometric interpretation of this algorithm, similar to Householder rotations~\cite{Householder58,Ajoy12} is described in \zfr{circle}. The required sample can be represented as a phasor on a circle, at an angle $\pi\zt(\R{sample})$. The algorithm is composed of two steps -- \I{propagate}, or \I{propagate and reflect}, associated with the application $U_0$ or $U_1$ operators respectively. Geometrically, one keeps propagating along points on the circle that differ by the required sample, and reflect every time when one trespasses into the left half circle (shaded region in \zfr{circle}). The algorithm ends when the phasor returns to the starting point. It is evident then for a sample $\xD\qt/N$, one needs $N$ operators in the construction.

As the optimal construction compensate the error at each step, it is significantly better  than the naive construction  $U_0^pU_1^q$ that lets the error accumulate (see \zfr{optimal-construction-graph}). Consequently, the number of supersamples achievable  via quantum interpolation scale linearly with the number of pulses to a large extent (\zfr{scaling}). 
More interestingly, this also implies that the error of all supersamples is approximately the same and bounded by the error of the $U_{1/2}$ as we shall show below. 

\subsubsection{Error of the Optimal Quantum Interpolation Construction}
While in principle we expect that each quantum interpolation construction, achieving supersampling at a different sampling point, might have a different error, here we show that for the same number of pulses, the error is always bounded by the error of  $U_{1/2}^N$ (see inset of \zfr{optimal-construction-graph}). 
For clarity, let us first consider the simple case when $N=2^k$ and calculate the error of any of the supersamples.  From the geometric picture in \zfr{circle}, all samples of the form $\ell \xD\qt/2^k$ for integral $(\ell,k)$ traverse the \I{same} set of points on the right half circle. 
Since $\xD a$ and $\xD b$ are now constrained to be points on the right half circle in \zfr{circle} the net error can be calculated from \zr{trap-error2} to be
\bea
\xe &\leq& 2\xD\qt\zt 2(\R{sum of all points on right half circle}) \non\\
&=& 2\xD\qt\zt 2\lb 2\sum_{\ell=0}^{2^{(k-1)}-1}\fr{\ell}{2^k} + \fr{1}{2}\rb=  2^k(2\xD\qt),
\eea
Hence the net error is bounded by $\xe\leq 2^{k-1}(4\xD\qt)$, exactly the error of $U^{2^{k}}_{1/2}$, which is the construction of the half-way-sample with the same number of pulses.  An analogous calculation and graphical approach can be made for general $N$, and once again it is easy to show that that the error of all supersamples is bounded by that of $U_{1/2}$.    This is  convenient because it allows a simplification of the error analysis of supersampling, which is bounded by the analytical results obtained in  \zsr{Uhalf}, where we evaluated the error of $U_{1/2}$  and quantified its dependence on the size of the sampling interval $\xD\qt$.

\subsection{Q-value as figure of merit}
\label{sec:Q-value}
\subsubsection{Gains in Q-value via quantum interpolation}
In order to characterize the boost in resolution granted by  supersampling, we introduce as figure of merit the  Q-value of the sensing peak, $Q=f_{\R{AC}} /\xD f_{\R{AC}}$. Given that the sensing peak arises at time $\qt=1/f_{\R{AC}}$, and the sensing linewidth in time units is $w$, we have  the frequency linewdith
\beq
\xD f = \fr{1}{\qt-w} - \fr{1}{\qt+w} \app \fr{2w}{\qt^2} = 2w f^2_{\R{AC}}
\eeq
giving $Q\approx1/\lsb 2w f_{\R{AC}}\rsb$. 
This definition of $Q$ describes the ability to resolve sensing peaks at different frequencies, where the minimum condition to resolve two  sensing peaks is that the peaks are separated by at least $2w$. 
The Q-value scales linearly with the number of pulses $N$, since ideally $w\propto1/N$. However, in practice $w$ is bounded by the finite timing resolution $\xD\qt$. Let us now evaluate the maximum achievable Q-value in this situation of limited $\xD\qt$. Then, the smallest linewidth one can achieve for sensing is $w=\xD\qt$, giving the bare Q-value
\beq
Q_{\R{bare}} = \fr{1}{f_{\R{AC}}(2\xD\qt)}.
\eeq
Quantum interpolation can boost the Q-value by overcoming the limits in sampling time, $\xD\qt$. This is shown for instance  in Fig. 2 of the main paper, where $\xD\qt=1$ns restricts $Q_{\R{bare}}$ to a maximum of 100. Thanks to quantum interpolation, where  the number of possible supersamples scales linearly with the pulse number, the Q-value is limited in principle only by the intrinsic linewidith of the sensing peak, given by the total time of the experiment. At the maximum allowed time, that is $T=T_2\app 1$ms,  we can estimate the Q-value under quantum interpolation using \zr{linewidth-time},
\beq
Q_{\R{supersample}} =\fr{\sq{2}\cos(\xa_j/2)}{\sin \lb\fr{2\pi^2}{T_2\xo_L}\rb}.
\eeq
Hence, due to quantum interpolation, one achieves a boost in the sensing $Q$ value by an amount,
\beq
Q_{\R{boost}} = \fr{Q_{\R{supersample}}}{Q_{\R{bare}}} =\fr{\xD\qt}{w} = \fr{\sq{2}\xD\qt\xo_L\cos(\xa_j/2)}{\sin \lb\fr{2\pi^2}{T_2\xo_L}\rb}
\eeq
where we have used the expression in \zr{subsampling3}.  Figure 2\textbf{(D)} of the main paper, obtained from the experiments in Fig. 2\textbf{(C)}, illustrates that the effective sensing $Q=f/\xD f$ can be boosted by a factor of 1000. However this experiment was performed with a total time of 115.2$\mu$s. At the $T_2$ of 1ms, one expects that $Q=8680$, i.e. one approach a Q-value of $10^4$ that could allow NV based sensors to measure fields from varied sources.

\subsubsection{Applications to quantum metrology experiments}
The Q-value provides a convenient measure to characterize the gains due to quantum interpolation. In our experiments, we were able to achieve substantial gains in Q-value over the bare limit set by the hardware. More broadly, quantum interpolation is useful for NV-based sensing of signals that have high Q,   for instance, signals with extremely small $\xD f$ (narrow linewidths or frequency differences) such as nuclear spins and chemical shifts, or signals with  high  frequency $f$, such as spin wave modes in ferromagnets. Many of these signals are currently out of reach because of the severe constraint set by timing resolution. 
 In the case of high frequency signals (for instance FMR in ferromagnets), we envision  sensing  higher harmonic of the signal for which the time $\qt$ is greater than the pulse width. Then, thanks to quantum interpolation the  time can be swept with a very fine step, allowing one to detect high-Q, high $f$ signals. 
 Quantum interpolation would then significantly broaden the impact  of  NV center as a probe for condensed matter systems, as we show in detail \ztr{table2} and \zfr{scatter}, where the Q-values are plotted against frequency.

\begin{figure*}
\begin{minipage}{.87\textwidth}
\begin{lstlisting}[caption = {Matlab code for the optimal quantum interpolation sequence construction}]
%% Basic blocks of delay time (separation between neighbouring pi pulses).

U_0 = [tau, 2*tau, tau];
U_1 = [tau+delta_tau, 2*(tau+delta_tau), tau+delta_tau];
% tau is the parameter we sweep in experiment and delta_tau is the smallest time step we can sweep by

%% Assemble the delay time in the order as in the optimal construction of a supersampling sequence.

delay={}; % Initialize cell array 'delay'
count=0; m=0;

for  j = 1: 4*n  %  Here n is XY8 cycle number
       m = m + sample;
   % sample is a fraction number between 0 and 1, rounded to multiples of 1/(4*n);
      if abs(m) <= 1/2
          count = count+1;
          delay{count} = U_0;
      else
          count  = count+1;
          delay{count} = U_1;
          m = m-1;
      end
end

% For instance, for n=1 and sample =1/4, one gets a delay cell array: delay={U_0 U_0 U_1 U_0}.

% Let us take as given functions which add a pulse or delay to a sequence.
% For simplicity, the carrier power and frequency are not shown as inputs and only phase is emphasized here.

function add_pi/2_pulse(phase)
function add_pi_pulse(phase)
function add_delay(delay_time)

%% Create a supersampling XY8-n sequence, combining pi pulses with correct phase and delay.

add_pi/2_pulse(0);

for j=1:4*n
 if mod(j,4)==1 || mod(j,4)==2
      add_delay(delay{j}(1,1));
      add_pi_pulse(0);  % pi|x
      add_delay(delay{j}(1,2));
      add_pi_pulse(90); % pi|y
      add_delay(delay{j}(1,3));
else
      add_delay(delay{j}(1,1));
      add_pi_pulse(90);  % pi|y
      add_delay(delay{j}(1,2));
      add_pi_pulse(0); % pi|x
      add_delay(delay{j}(1,3));
end
end

add_pi/2_pulse(0);
\end{lstlisting}
\end{minipage}
\end{figure*}

\subsection{Code for the construction of the optimal supersampling sequence}
\label{sec:code}
Here we present a simple code  (in MATLAB) that allows  constructing  the optimal interpolation sequence for a desired sampling time $q/(p+q)\xD\qt$ (see \zfr{error-matrix}). 
The algorithm yields an array of time delays for how the  basic CPMG building blocks $U_0$ and $U_1$ (each consisting of 3 rotations) should be ordered and the  $\pi$-pulses  phases  chosen according to the XY8 scheme.

%

\end{document}